\documentclass[12pt]{article}
\usepackage{jheppub}
\pdfoutput=1
\usepackage{epstopdf}
\usepackage{slashed}
\usepackage{mathrsfs}
\usepackage{psfrag}
\usepackage{epsf}
\usepackage{graphicx,epsfig}
\usepackage{amssymb}
\usepackage{epstopdf}
\usepackage{float}
\usepackage{braket}
\usepackage{hyperref}
\usepackage{xcolor}
\usepackage{mathtools}
\usepackage{physics}
\usepackage{tikz-feynman}
\tikzfeynmanset{compat=1.0.0}
\usepackage[export]{adjustbox}
\usepackage{relsize}
\newcommand{\eqs}[2][0.3]{\includegraphics[width=#1\linewidth, valign=c]{#2}}

\usepackage{amsmath,array,amsfonts,wrapfig,lscape,multirow,longtable,rotating,subfigure,makecell}
\usepackage{url}

\newcolumntype{C}[1]{>{\centering\arraybackslash}m{#1}}

\newcommand{\rayan}[1]{{\bf [RH: #1]}}

\newcommand{\K}[1]{k_{\perp #1}}
\newcommand{\Kb}[1]{\kappa_{\perp #1}}
\newcommand{\KT}[1]{\tilde{k}_{\perp #1}}
\newcommand{\KTb}[1]{\tilde{\kappa}_{\perp #1}}
\newcommand{\pM}[1]{\widetilde{P}^{#1}}
\newcommand{\pN}[1]{\widetilde{P}'^{#1}}

\def\beq{\begin{equation}}
\def\eeq{\end{equation}}

\newcolumntype{C}[1]{>{\centering\arraybackslash}m{#1}}

\newcommand{\comments}[1]{}

\def\cM{{\cal M}}  
  
 \def\cT{{\cal T}}

\def\eps{\epsilon}

\graphicspath{{./images/}}

\def\beq{\begin{equation}}
\def\eeq{\end{equation}}
\def\bsp#1\esp{\begin{split}#1\end{split}}
\def\cM{{\cal M}}

\preprint{CERN-TH-2020-108, CP3-20-31}

\title{Tree-level splitting amplitudes for a gluon into four collinear partons}
\author[a]{{\small Vittorio Del Duca}\footnote{On leave from INFN, Laboratori Nazionali di Frascati, Italy.}}
\author[b]{{\small Claude Duhr}}
\author[a]{{\small Rayan Haindl}}
\author[a]{{\small Achilleas Lazopoulos}}
\author[c]{{\small Martin Michel}}

\affiliation[a]{Institute for Theoretical Physics, ETH Z\"{u}rich, 8093 Z\"{u}rich, Switzerland.}
\affiliation[b]{Theoretical Physics Department, CERN, CH-1211 Geneva 23, Switzerland.}
\affiliation[c]{Center for Cosmology, Particle Physics and Phenomenology (CP3),\\
UCLouvain, B-1348, Louvain-La-Neuve, Belgium.}

\emailAdd{delducav@itp.phys.ethz.ch, claude.duhr@cern.ch, haindlr@phys.ethz.ch, lazopoli@phys.ethz.ch, martin.michel@uclouvain.be}

\abstract{
We compute in conventional dimensional regularisation
the tree-level splitting amplitudes for a gluon parent which splits into four collinear partons. This is part of the universal infrared behaviour of the QCD scattering amplitudes at 
next-to-next-to-next-to-leading order (${\rm N^3LO}$) in the strong coupling constant. Combined with our earlier results for a quark parent, this completes the set of tree-level splitting amplitudes required at this order.
We also study iterated collinear limits where a subset of the four collinear partons become themselves collinear. 
}
\begin{document}

\maketitle

\section{Introduction}
\label{sec:intro}

With the discovery of the Higgs boson by the Large Hadron Collider (LHC) at CERN~\cite{Aad:2012tfa,Chatrchyan:2012xdj}, the Standard Model (SM) of particle physics is a complete theory without any free parameters. Current and future collider experiments will be able to test the SM with an increasing level of precision. 
The requested precision poses a severe challenge to theory and calls for the development of improved techniques for theoretical predictions accurate at the percent level. One of the main approaches to theoretical collider phenomenology is perturbation theory, in which observables are expanded in the small coupling constants of the theory. Here we focus on QCD observables, and consequently we will be concerned with the computation of higher orders in the strong coupling constant $\alpha_s$ of QCD. 

Often, leading order predictions in QCD are not reliable. For example it is known that in the case of Higgs production the next-to-leading order corrections almost double the value of the cross section~\cite{Graudenz:1992pv,Spira:1995rr}.
The next-to-next-to-leading order (NNLO) corrections (in the limit where the top-quark is infinitely heavy) further increase the cross section, and it is only after the inclusion of the next-to-next-to-next-to-leading order (N$^3$LO) corrections that a reliable estimate of the cross section with a residual uncertainty of only a few percent is obtained~\cite{Anastasiou:2015ema,Anastasiou:2016cez,Mistlberger:2018etf}.
This example illustrates that reliable QCD predictions at the percent level can most likely only be achieved after the inclusion of N$^3$LO corrections. So far, however, only very few hadron collider processes are known at this order in perturbation theory. Inclusive N$^3$LO cross sections are known for Higgs in gluon fusion~\cite{Anastasiou:2015ema,Anastasiou:2016cez,Mistlberger:2018etf}, bottom-quark fusion~\cite{Duhr:2019kwi,Duhr:2020kzd} and vector boson fusion in the DIS approach~\cite{Dreyer:2016oyx,Dreyer:2018qbw}, as well as double Higgs production~\cite{Chen:2019lzz,Chen:2019fhs} and Drell-Yan production (via the intermediate of an off-shell photon)~\cite{Duhr:2020seh}. At the differential level, only the Higgs rapidity and transverse momentum distributions in vector boson fusion in the DIS approximation~\cite{Cieri:2018oms,Dulat:2018bfe, Dreyer:2016oyx} are known.

One of the reasons progress towards more results at N$^3$LO is difficult lies in the fact that an observable at N$^k$LO receives contributions from processes with up to $k$ additional partons in the final state. Each such contribution is individually infrared divergent, with divergences arising in particular from the integration over regions of phase space where the emitted partons are either soft or collinear. Several techniques have been developed at NNLO to compute the relevant phase space integrals with generic acceptance cuts~\cite{Binoth:2000ps,Anastasiou:2005cb,Catani:2007vq,Boughezal:2015eha,Gaunt:2015pea,
GehrmannDeRidder:2005cm,Daleo:2006xa,GehrmannDeRidder:2005aw,
GehrmannDeRidder:2005hi,Daleo:2009yj,Gehrmann:2011wi,Boughezal:2010mc,GehrmannDeRidder:2012ja,Currie:2013vh,
Somogyi:2005xz,Somogyi:2006da,
Somogyi:2006db,Somogyi:2008fc,Aglietti:2008fe,Somogyi:2009ri,
Bolzoni:2009ye,Bolzoni:2010bt,DelDuca:2013kw,Somogyi:2013yk, 
Czakon:2010td,Czakon:2011ve,Czakon:2014oma,Czakon:2019tmo, 
Caola:2017dug,Caola:2018pxp,Delto:2019asp,Caola:2019nzf,Caola:2019pfz,Cacciari:2015jma}. With some abuse of language, we refer to these techniques  in the following collectively as \emph{subtraction methods}. Developing  subtracting methods beyond NNLO will be an important step towards obtaining more predictions with percent accuracy.\footnote{Some of these methods have already been successfully applied at N$^3$LO for processes with a simple final state structure~\cite{Dreyer:2016oyx,Currie:2018fgr,Dreyer:2018qbw,Cieri:2018oms}.}

An important ingredient in the development of subtraction methods  is the universal behaviour of the QCD scattering amplitudes in the infrared limits, embodied in universal soft and collinear currents. At NNLO these currents have been computed more than a decade ago, and they include splitting amplitudes for three partons at tree-level and two partons at one-loop, as well as soft currents for the emission of two soft partons at tree-level and one soft parton at one-loop~\cite{Bern:1994zx,Campbell:1997hg,Catani:1999ss,DelDuca:1999iql,Bern:1998sc,Kosower:1999rx,Bern:1999ry,Catani:2000pi,Kosower:2002su}. Over the last few years, results have also become available for the universal currents at N$^3$LO. The two-loop currents for the emission of two collinear or one soft parton were obtained in refs.~\cite{Bern:2004cz,Badger:2004uk,Duhr:2014nda,Duhr:2013msa,Li:2013lsa,Dixon:2019lnw}, and the one-loop current for the emission of three collinear partons is given in refs.~\cite{Catani:2003vu,Badger:2015cxa}.\footnote{One-loop currents with three collinear partons are also known for
mixed QCD+QED cases~\cite{Sborlini:2014mpa}.} 
The tree-level currents include the soft current for the emission of three soft partons~\cite{Catani:2019nqv} and the splitting amplitudes for four collinear partons. The latter have been obtained at the amplitude-level in four dimensions in refs.~\cite{DelDuca:1999iql,Birthwright:2005ak,Duhr:2006}. Recently, we have published the splitting amplitudes for the squared matrix element in dimensional regularisation, in the case where the parent parton is a quark~\cite{DelDuca:2019ggv}. The main purpose of this paper is to complete the set of tree-level splitting amplitudes at N$^3$LO by providing analytic results in dimensional regularisation for a gluon splitting into four partons.

This paper is organised as follows:
In Sec.~\ref{sec:setup}, we review the collinear limit of tree-level amplitudes and give a precise definition of the quantities that we want to compute. 
In Sec.~\ref{sec:collinear}, we present the main result of our paper, namely the computation of the tree-level splitting amplitudes for a gluon parent to split into four collinear partons. The explicit results are too long to be recorded in this paper and are made available in computer-readable form~\cite{QuadColKernelsWebsite}.  
In Sec.~\ref{sec:nested_collinear} we study the collinear limit of the splitting amplitudes themselves, and we define new universal objects which appear in these iterated limits. We include several appendices with technical material omitted throughout the main text.

%%%%%%%%%%%%%%%%%%%%%%%%%%
\section{Multiple collinear limits}
\label{sec:setup}

We examine the behaviour of tree-level QCD amplitudes in the limit where a given number of massless partons become collinear. Namely, we consider the scattering of $n$ massless particles with momenta $p_i$ and with flavour, spin and colour quantum numbers $f_i$, $s_i$ and $c_i$, respectively, and we analyse the behaviour of the amplitude as $m$ partons of momenta $p_1,\ldots,p_m$ become simultaneously collinear to some light-like direction $\pM{}$. In this limit, the leading behaviour is described by the amplitude for the production of a massless particle of momentum $\pM{}$ from a scattering of the particles that do not take part in the collinear limit, multiplied by a universal factor, termed the \emph{splitting amplitude}, which depends only on the $m$ partons in the collinear set.

In order to parametrise the approach to the collinear limit, we introduce a light-cone decomposition for all the momenta in the $m$-parton collinear set,
\begin{equation} 
\label{eq:collinear_momenta}
p_i^\mu = x_i\pM{\mu} + \K{i}^\mu - \frac{\K{i}^2}{2 x_i}\frac{n^\mu}{\pM{}\cdot n}\,,
\qquad
i = 1,\ldots, m\,,
\end{equation}
where the light-like momentum $\pM{}$ specifies the collinear direction, 
$\pM{} \cdot k_{\perp i} = 0$, $x_i$ are the longitudinal momentum fractions with respect to the parent momentum $P^\mu = \sum_{i=1}^m p_i^\mu$ and
$n^\mu$ is an auxiliary light-like vector such that $n\cdot\K{i} = 0$ and $n\cdot p_{i} \neq 0\neq n\cdot\pM{}$, and which specifies how the collinear direction is approached. The collinear limit is then defined as the limit in which the transverse momenta $\K{i}$ approach zero at the same rate. This definition of the collinear limit is frame-independent, and it only depends on the collinear direction $\pM{}$ and the transverse momenta $\K{i}$. In particular it is independent of the choice of the auxiliary vector $n$.

The variables that appear in eq.~\eqref{eq:collinear_momenta} are unconstrained apart from on-shellness and transversality, $n\cdot \K{i} = \pM{}\cdot \K{i} = 0$, and so the sums of the momentum fractions $x_i$ and the transverse momenta $\K{i}$ are unconstrained.
%\beq
%\sum_{i=1}^m x_i \neq 1 {\rm~~and~~} \sum_{i=1}^mk_{\perp i}\neq 0\,.
%\eeq
%Therefore, the parametrisation in eq.~\eqref{eq:collinear_momenta} seems to depend on $(D-1)m+2$ degrees of freedom in $D$ space-time dimensions: the $m$ variables $x_i$ and $\K{i}$ in addition to the two light-like directions $\pM{}$ and $n$. This naive counting seems to be at odds with the fact that a set of $m$ light-like momenta (that do not sum up to zero) depend on $(D-1)m$ degrees of freedom. This apparent conundrum is resolved upon noting that 
However, the collinear limit is invariant under longitudinal boosts in the direction of the parent momentum $P = \sum_{i =1}^m p_i$. We trade $x_i$ and $\K{i}$ for new quantities $z_i$ and $\KT{i}$ that are boost-invariant in the direction of the parent momentum. In refs.~\cite{Catani:1999ss,DelDuca:2019ggv}, it was shown that a convenient set of such variables is given by
\begin{equation}
z_i = \frac{x_i}{\sum_{j=1}^m x_j} = \frac{p_i\cdot n}{P\cdot n}\,, 
\qquad \KT{i}^\mu = \K{i}^{\mu}- z_i\sum_{j=1}^m \K{j}^{\mu}\,,
\qquad
i = 1,\ldots, m\,.
\label{eq:boost}
\end{equation}
It is easy to see that these new variables satisfy the constraints,
\beq
\sum_{i=1}^m z_i =1 {\rm~~and~~} \sum_{i=1}^m\KT{i}^\mu = 0\,.
\label{eq:constraint}
\eeq
From now on, we only work with these variables, and in order to avoid cluttering notation, we shall drop the tilde on the transverse momenta.
%thereby reducing the number of degrees of freedom to $(D-1)m$.

In the limit where a subset of massless particles is collinear, a scattering amplitude factorises as~\cite{Amati:1978wx,Amati:1978by,Ellis:1978sf}
\beq\bsp\label{eq:split_def}
\mathscr{C}_{1\ldots m}\cM&_{f_1\ldots f_n}^{c_1\ldots c_n;s_1\ldots s_n}(p_1,\ldots,p_n)\\
&\, = {\bf Sp}_{ff_1\ldots f_m}^{c,c_1\ldots c_m;s,s_1\ldots s_m}\,\cM_{ff_{m+1}\ldots f_n}^{c,c_{m+1}\ldots c_n;s,s_{m+1}\ldots s_n}(\pM{},p_{m+1},\ldots,p_n)\,,
\esp\eeq
where $\mathscr{C}_{1\ldots m}$ indicates that the equality holds up to terms that are power-suppressed in the collinear limit, while $f$, $s$ and $c$ respectively denote the flavour, spin and colour indices of the parent particle. 
The quantity ${\bf Sp}$ appearing on the right-hand side is the splitting amplitude, which depends only on the kinematics and the quantum numbers in the collinear set. 

For an amplitude whose collinear massless particles occur all in the final state,
the factorisation in eq.~\eqref{eq:split_def} is valid to all orders in perturbation theory.\footnote{When the subset of collinear particles contains also initial-state particles, the factorisation in eq.~\eqref{eq:split_def} is valid in general only for tree amplitudes~\cite{Catani:2011st}.}
Accordingly, also the squared matrix element factorises, 
\beq
\left|\cM_{f_1\ldots f_n}(p_1,\ldots,p_n)\right|^2 \equiv \sum_{\substack{(s_1,\ldots,s_n)\\(c_1,\ldots,c_n)}}\left|\cM_{f_1\ldots f_n}^{c_1\ldots c_n;s_1\ldots s_n}(p_1,\ldots,p_n)\right|^2\,,
\eeq
where in the short-hand notation of the left-hand side
the sum over all spin and colour indices of the matrix element
is understood. The factorisation of the squared matrix element can be written as
\beq
\mathscr{C}_{1\ldots m}\left|\cM_{f_1\ldots f_n}(p_1,\ldots,p_n)\right|^2 = \Bigg(\frac{2\mu^{2\epsilon}\,g_s^2}{s_{1\ldots m}}\Bigg)^{m-1} \,\hat{P}^{ss'}_{f_1\ldots f_m}\, \mathcal{T}_{ff_{m+1}\ldots f_n}^{ss'}(\pM{},p_{m+1},\ldots,p_n)\,,
\label{eq:factorisation}
\eeq
where $g_s$ is the strong coupling constant and $\mu$ is the scale introduced by dimensional regularisation, and a sum over
repeated indices, in this case the spin indices $s$ and $s'$, is implicit, and we introduced the Mandelstam invariant,
\beq
s_{1\ldots m} \equiv (p_1+\ldots+p_m)^2\,.
\eeq
$\mathcal{T}_{ff_{m+1}\ldots f_n}^{ss'}$ denotes the helicity tensor obtained by not summing over the spin indices of the parent parton,
\beq\label{eq:T_def}
 \mathcal{T}_{ff_{m+1}\ldots f_n}^{ss'} \equiv \sum_{\substack{(s_{m+1},\ldots,s_n)\\(c,c_{m+1},\ldots,c_n)}}\cM_{ff_{m+1}\ldots f_n}^{c,c_{m+1}\ldots c_n;s,s_{m+1}\ldots s_n}\,\left[\cM_{ff_{m+1}\ldots f_n}^{c,c_{m+1}\ldots c_n;s',s_{m+1}\ldots s_n}\right]^{\ast}\,,
\eeq
where for brevity we have suppressed the momenta on which the amplitude depends. The tensorial structure of the factorisation in eq.~\eqref{eq:factorisation} is necessary to correctly capture all spin correlations. Due to colour conservation in the hard amplitude there are no non-trivial colour correlations, and we therefore sum over the colour $c$ of the parent parton in eq.~\eqref{eq:T_def}. The quantity $\hat{P}^{ss'}_{f_1\ldots f_m}$ in eq.~\eqref{eq:factorisation} is the (polarised) splitting amplitude for the squared matrix element, which is related to ${\bf Sp}$ by 
\beq\label{eq:P_def}
\Bigg(\frac{2\mu^{2\epsilon}\,g_s^2}{s_{1\ldots m}}\Bigg)^{m-1} \,\hat{P}^{ss'}_{f_1\ldots f_m} =
\frac1{\mathcal{C}_{f}}
\sum_{\substack{(s_{1},\ldots,s_m)\\(c,c_{1},\ldots,c_m)}}  {\bf Sp}_{ff_1\ldots f_m}^{c,c_1\ldots c_m;s,s_1\ldots s_m}\,\left[{\bf Sp}_{ff_1\ldots f_m}^{c,c_1\ldots c_m;s',s_1\ldots s_m}\right]^{\ast}\,,
\eeq
where $\mathcal{C}_{f}$ is the number of colour degrees of freedom of the parent parton with flavour $f$, i.e., $\mathcal{C}_{g}=N_c^2-1$ for a gluon and $\mathcal{C}_q=N_c$ for a quark. In eqs.~(\ref{eq:factorisation}), (\ref{eq:P_def}) and henceforth,
the dependence of the splitting amplitude
on the transverse momenta $\KT{i}$ and momentum fractions $z_i$ of the particles in the collinear set is understood.
Further, in QCD the flavour of the parent is uniquely determined by the flavours of the particles in the collinear set, thus we suppress the dependence of the splitting amplitude on the left-hand-side of eq.~\eqref{eq:P_def} on the flavour of the parent parton.

Splitting amplitudes for the squared matrix element have been computed at tree level for the emission of up to three collinear partons in refs.~\cite{Campbell:1997hg,Catani:1999ss},
and for the emission of four collinear partons
out of a parent quark in ref.~\cite{DelDuca:2019ggv}.
The goal of this paper is to compute the tree-level splitting amplitudes for the squared matrix element for the emission of up to four partons out of a parent gluon, thus completing the set of splitting amplitudes for the emission of up to four collinear partons in QCD.\footnote{As in
ref.~\cite{DelDuca:2019ggv}, we refer to both ${\bf Sp}$ and $\hat{P}$ simply as splitting amplitudes.}

%%%%%%%%%%%%%%%%%%%%%%%%%%%%%%%%%%%%%%
\section{Gluon-initiated splitting amplitudes}
\label{sec:collinear}
In this section we present the computation of the gluon-initiated tree-level splitting amplitudes for $m=4$ collinear partons,\footnote{The constraints in eq.~(\ref{eq:constraint})
have not been imposed on our results. This may allow us, through crossing symmetry, to readily 
obtain the splitting amplitudes for initial-state collinear emissions~\cite{deFlorian:2001zd}.} which is the main result of our paper. The computation follows the same lines as that for $m=3$ collinear partons in ref.~\cite{Catani:1999ss}. Our results for the splitting amplitudes are too lengthy to be presented in printed form, but we make them available in computer-readable form~\cite{QuadColKernelsWebsite}. 

%%%%%%%%%%%%%%%% COMMENT OUT %%%%%%%%%%%%%%%%%%%%
{
}
%%%%%%%%%%%%%%%% END COMMENT OUT %%%%%%%%%%%%%%%%%%%%

%%%%%%%%%%%%%%%% COMMENT OUT %%%%%%%%%%%%%%%%%%%%
{
}
%%%%%%%%%%%%%%%% END COMMENT OUT %%%%%%%%%%%%%%%%%%%%

In order to compute an $m$-parton splitting amplitude, we start from an on-shell amplitude for $n=m+3$ partons and take $m$ of them collinear. We perform a uniform rescaling of the transverse momenta $k_{\perp i}$ in eq.~\eqref{eq:collinear_momenta} by a small parameter $\lambda$,
\begin{align}
k_{\perp i} \to \lambda\, k_{\perp i}\,, \quad 1\le i\le m\,.
\end{align}
This ensures that in the collinear limit $\lambda \to 0$ the $k_{\perp i}$ approach zero at the same rate. We then expand the matrix element into a Laurent series around $\lambda=0$. The leading term corresponds to the coefficient of $1/\lambda^{2(m-1)}$, which is universal and is  described by the collinear factorisation in eq.~\eqref{eq:factorisation}. 

While the final result of this operation is of course gauge independent, the set of Feynman diagrams that contribute to the leading behaviour in $\lambda$ depend on the gauge choice. We would therefore like to choose a gauge that simplifies the computation as much as possible, e.g., by minimising the number of (interfering) Feynman diagrams that contribute in the collinear limit. 
In ref.~\cite{Catani:1999ss} it was argued that it is convenient to work in a physical gauge (e.g., axial gauge), because contributions from Feynman diagrams where collinear partons are separated by a hard propagator are subleading in the collinear limit. 
Here we work in axial gauge, where the gluon field is subject to the following conditions,
\beq
\partial_{\mu}A^{\mu} = n_{\mu}A^{\mu} = 0\,,
\eeq
where $n$ is an arbitrary light-like reference vector. In this gauge, the gluon propagator takes the form,
\beq\label{eq:axial_propagator}
\eqs[0.2]{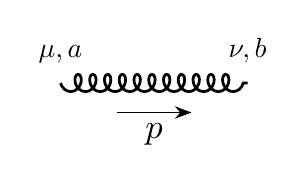}=\hspace{0.3cm}
\frac{i\,\delta^{ab}\,d^{\mu\nu}(p,n)}{p^2+i\varepsilon}\,,\qquad d^{\mu\nu}(p,n) = -g^{\mu\nu}+\frac{p^\mu n^\nu+n^\mu p^{\nu}}{p\cdot n}\,.
\eeq
For $p^2=0$, the polarisation tensor $d^{\mu\nu}(p,n)$ can be interpreted in three different ways: First, it is the projector onto the $(D-2)$-dimensional space transverse to $p$ and $n$. Second, it is the metric tensor induced by the $D$-dimensional Minkowski metric on that space. Finally, it is the sum of all physical polarisations of a gluon with momentum $p$,
\beq\label{eq:polarisation_sum}
\sum_{s}\varepsilon_s^{\mu}(p,n)\varepsilon_s^{\nu}(p,n) = d^{\mu\nu}(p,n)\,,
\eeq
where $\varepsilon_s^{\mu}(p,n)$ is the polarisation vector for a gluon with momentum $p$ and transverse polarisation $s=1,\ldots, (D-2)$.
In principle, we may choose a different reference vector for each gluon, as long as it is not orthogonal to the momentum. In our case, it is convenient to choose all gauge reference vectors to coincide with the reference vector $n$ appearing in the definition of the collinear limit in eq.~\eqref{eq:collinear_momenta}. We can then write the collinear factorisation for a parent gluon in terms of Lorentz indices rather than helicities~\cite{Catani:1999ss},
\beq
\mathscr{C}_{1\ldots m}\left|\cM_{f_1\ldots f_n}(p_1,\ldots,p_n)\right|^2 = \Bigg(\frac{2\mu^{2\epsilon}\,g_s^2}{s_{1\ldots m}}\Bigg)^{m-1} \,\hat{P}^{\mu\nu}_{f_1\ldots f_m}\, \mathcal{T}_{ff_{m+1}\ldots f_n,\mu\nu}(\pM{},p_{m+1},\ldots,p_n)\,,
\label{eq:gluon_factorisation}
\eeq
where quantities with open Lorentz indices are obtained by amputating the polarisation vectors and inserting the polarisation sum in eq.~\eqref{eq:polarisation_sum}. Note that in passing from eq.~\eqref{eq:factorisation} to eq.~\eqref{eq:gluon_factorisation} we have implicitly used gauge invariance to eliminate the gauge dependent terms in eq.~\eqref{eq:polarisation_sum}.
Indeed, since physical polarisation states are transverse, only the transverse part of a Lorentz tensor carries physical information. This is because the non-transverse part vanishes upon contraction with a physical polarisation vector. As a consequence, the helicity tensor $\mathcal{T}_{ff_{m+1}\ldots f_n}^{\mu\nu}$ in eq.~\eqref{eq:gluon_factorisation} can be chosen to satisfy the transversality condition,
\beq
\pM{}_\mu\,\mathcal{T}_{ff_{m+1}\ldots f_n}^{\mu\nu} = \pM{}_\nu\,\mathcal{T}_{ff_{m+1}\ldots f_n}^{\mu\nu} = 0\,.
\label{eq:transverse_T}
\eeq
With this choice, the complete tensor structure of the splitting amplitude contains terms involving the transverse momenta of the collinear partons~\cite{Catani:1999ss},
\beq\label{eq:P_tensor}
\hat{P}^{\mu\nu}_{f_1\ldots f_m} = g^{\mu\nu}\,A^{(g)}_{f_1\ldots f_m} + \sum_{i,j=1}^m\,\frac{\tilde{k}_{\perp i}^{\mu}\tilde{k}_{\perp j}^{\nu}}{s_{1\ldots m}}\,B^{(g)}_{ij,f_1\ldots f_m}\,.
\eeq
We stress, however, that the splitting amplitude defined in this way does not vanish upon contraction with $\pM{}$.
%The gluon-initiated splitting amplitudes for $m=2$ and $m=3$ read~\cite{XX},
%\beq\bsp
%\langle \hat{P}_{gg}\rangle &\,=\ldots\,,\\
%\langle \hat{P}_{q\bar{q}}\rangle &\,=\ldots\,,\\
%\langle \hat{P}_{ggg}\rangle &\,=\ldots\,,\\
%\langle \hat{P}_{gq\bar{q}}\rangle &\,=\ldots\,.
%\esp\eeq
%\fi
%Let us make a comment about gauge invariance. 

Since we work in axial gauge, we do not consider the  subset of Feynman diagrams where collinear partons are separated by a hard propagator. The sum of all relevant diagrams can be cast in the form,
\beq\bsp
\mathscr{C}_{1\ldots m}&\left|\cM_{f_1\ldots f_n}(p_1,\ldots,p_n)\right|^2 = \\
&\mathscr{C}_{1\ldots m}\left[
 \Bigg(\frac{2\mu^{2\epsilon}g_s^2}{s_{1\ldots m}}\Bigg)^{m-1} 
\left[ \mathcal{M}^{(n)\, s}_{ff_{m+1}\ldots f_n}\right]^\ast V^{(n)\, ss'}_{f_1\ldots f_m}(p_1,\ldots,p_m) \mathcal{M}^{(n)\, s'}_{ff_{m+1}\ldots f_n}\right]\,,
\label{eq:V_fac2}
\esp\eeq
where a sum over the spin indices $s$, $s'$ of the intermediate state is understood, and we suppress all colour and spin indices of the external partons. Here $\mathcal{M}^{(n)\, s}_{ff_{m+1}\ldots f_n} \equiv \mathcal{M}^{(n)\, s}_{ff_{m+1}\ldots f_n}(P,p_{m+1},\ldots,p_n)$ denotes the sum of all Feynman diagrams with an off-shell leg with momentum $P$, flavour $f$ and spin $s$. Note that this subset of Feynman diagrams is by itself not gauge invariant, and the superscript $(n)$ indicates the dependence on the reference vector. 
The squared off-shell current $ V^{(n)\, ss'}_{f_1\ldots f_m}$ may be written as the interference of two colour-dressed off-shell currents,
\begin{equation}\label{eq:V_to_j}
\!\!\Bigg(\frac{2\mu^{2\epsilon}g_s^2}{s_{1\ldots m}}\Bigg)^{m-1} V^{(n)\, ss'}_{f_1\ldots f_m}(p_1,\ldots,p_m) =
\frac1{\mathcal{C}_{f}} \sum_{\substack{(s_1,\ldots,s_m)\\(c,c_1,\ldots,c_m)}}\,
\left[{\mathrm J}_{f_1\ldots f_m}^{c,c_1\ldots c_m;s's_1\ldots s_m}\right]^{\ast}\,
{\mathrm J}_{f_1\ldots f_m}^{c,c_1\ldots c_m;ss_1\ldots s_m}\,,
\end{equation}
where $\mathcal{C}_{f}$ is defined after eq.~(\ref{eq:P_def}). Note that also $ V^{(n)\, ss'}_{f_1\ldots f_m}$ depends on the gauge vector $n$ before the collinear limit is taken. Since the collinear limit is gauge invariant, this dependence disappears in the limit, and the squared off-shell current reduces to the splitting amplitude,
\beq\label{eq:V_to_P}
\mathscr{C}_{1\ldots m}V^{(n)\, ss'}_{f_1\ldots f_m}(p_1,\ldots,p_m) = \hat{P}_{f_1\ldots f_m}^{ss'}\,.
\eeq

%Using eq.~\eqref{eq:V_to_P} we can substantially reduce the number of interfering Feynman diagrams that we need to evaluate. This strategy can always be used, independently of the flavour of the parent parton. We focus in this paper solely on the case where the parent parton is a gluon, $f=g$. 

We have computed all gluon-initiated splitting amplitudes up to $m=4$, and we reproduce all known results for the cases $m=2$ and 3. The results for $m=4$ are new and are presented for the first time in this paper. There are four different gluon-initiated splitting amplitudes,
\begin{align}
    g &\to \bar{q}'q'\bar{q}q\,, \qquad
    g \to \bar{q}q\bar qq\,, \qquad
    g \to \bar{q}ggq\,, \qquad
    g \to gggg\,.
\end{align}
In the remainder of this section we discuss in more detail the computation of these splitting amplitudes. The explicit results are available in computer-readable form~\cite{QuadColKernelsWebsite}.

%%%%%%%%%%%%%%%%%%%%%%%%%%%%%%%%%%%%%%%%%%%%%%%%%
%%%%%%%%%%%%%%% g -> q' q' q q %%%%%%%%%%%%%%%%%%

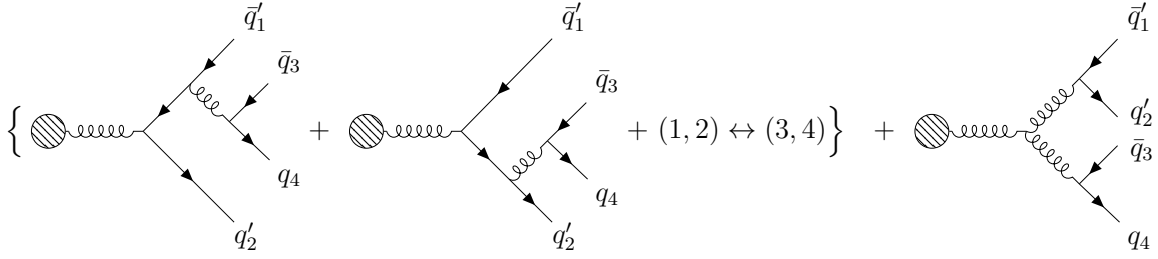
\begin{figure}[!t]
\centering
\begin{equation} \nonumber
\begin{split} 
&
\scalebox{0.9}{
\begin{tikzpicture}[baseline=(a)]
\begin{feynman}[inline=(a)]
\node[blob, scale=1] (blob1);
\vertex [left=0.5cm of blob1] (bracket){\(\Big\{\)};
\vertex [right=0.3cm of blob1] (a);
\vertex [above right = 0.3cm of a, draw=none] (null1){\(\)};
\vertex [right=1.1cm of a] (b);
\vertex [above right=1cm of b] (c);
\vertex [above right=0.9 cm of c] (q1){\(\bar{q}'_1\)};
\vertex [below right = 0.8cm of c] (gluon);
\vertex [below right=0.8cm of gluon] (q4){\(q_4\)};
\vertex [above right = 0.8cm of gluon] (q3){\(\bar{q}_3\)};
\vertex [below right=1.9cm of b] (q2);
\vertex [below right=1.7cm of b] (label) {\(q'_2\)};
\vertex [right=2.3cm of b, draw=none] (plus){\(+\)};
\diagram*{
(blob1) -- (a) -- [gluon] (b) -- [anti fermion] (c),
(c) -- [anti fermion] (q1),
(c) -- [gluon] (gluon)--[fermion] (q4),
(q3) -- [fermion] (gluon),
(b) -- [fermion] (q2)
};
\end{feynman}
\end{tikzpicture} 
\begin{tikzpicture}[baseline=(a)]
\begin{feynman}[inline=(a)]
\node[blob, scale=1] (blob1);
\vertex [right=0.3cm of blob1] (a);
\vertex [above right = 0.3cm of a, draw=none] (null1){\(\)};
\vertex [right=1.1cm of a] (b);
\vertex [above right=1.9 cm of b] (q1){\(\bar{q}'_1\)};
\vertex [below right=1cm of b] (c);
\vertex [above right = 0.8cm of c] (g);
\vertex [above right=0.8cm of g] (q3){\(\bar{q}_3\)};
\vertex [below right = 0.8cm of g] (q4){\(q_4\)};
\vertex [below right=0.9cm of c] (q2);
\vertex [below right=0.7cm of c] (label) {\(q'_2\)};
\vertex [right=2.3cm of b, draw=none] (plus){\(+\text{  }(1,2) \leftrightarrow (3,4)\Big\}\)};
\diagram*{
(blob1) -- (a) -- [gluon] (b) -- [anti fermion] (q1),
(b) -- [fermion] (c),
(c) -- [fermion] (q2),
(c) -- [gluon] (g)--[anti fermion] (q3),
(g) -- [fermion] (q4)
};
\end{feynman}
\end{tikzpicture} 
\begin{tikzpicture}[baseline=(a)]
\begin{feynman}[inline=(a)]
\node[blob, scale=1] (blob1);
\vertex [right=0.3cm of blob1] (a);
\vertex [left=0.7cm of blob1] (plus){\(+\)};
\vertex [below = 0.1cm of a, draw=none] (null1);
\vertex [right=1.1cm of a] (d1);
\vertex [above right = 1.1cm of d1] (g1);
\vertex [above right = 0.4cm of d1] (n1);
\vertex [right = 1.1cm of n1] (n2) {\(q'_2\)};
\vertex [below right = 1.1cm of d1] (g2);
\vertex [above right= 0.8cm of g1] (q1) {\(\bar{q}'_1\)};
\vertex [below right= 0.8cm of g1] (q2);
\vertex [below right = 0.4cm of d1] (m1);
\vertex [right = 1.1cm of m1] (m2) {\(\bar{q}_3\)};
\vertex [above right= 0.8cm of g2] (q3);
\vertex [below right=0.8cm of g2] (q4) {\(q_4\)};
\diagram*{
(blob1)-- (a) -- [gluon]  (d1) --  [gluon] (g1) -- [anti fermion] (q1), %momentum'={\small\(p_{1234}\)}
(g1) -- [fermion] (q2),
(d1) -- [gluon] (g2) -- [anti fermion] (q3),
(g2) -- [fermion] (q4)
};
\end{feynman}
\end{tikzpicture}}
\end{split}
\end{equation}
\caption{The diagrams contributing to the off-shell current $g \to \bar q_1^\prime q_2^\prime \bar q_3 q_4$. In the case of identical quarks we also need to include diagrams where the anti-quarks $q_1$ and $q_3$ are exchanged.}
\label{fig:q'q'qq}
\end{figure}

Let us start by discussing the simplest splitting process, the collinear decay $g \to \bar{q}' q' \bar q q$ with different quark flavours. There are five diagrams that contribute to the off-shell current ${\mathrm J}^\mu_{\bar{q}' q' \bar q q}$ in eq.~\eqref{eq:V_to_j}. The diagrams are shown in fig. \ref{fig:q'q'qq}.
Going through the steps outlined above, we find that the result for the splitting amplitude $g \to \bar{q}' q' \bar q q$ can be decomposed into an `abelian' and a `non-abelian' part,
\begin{equation}
\label{eq:Pq_qb'q'qbq}
\hat{P}^{\mu\nu}_{\bar{q}^\prime_1 q^\prime_2 \bar q_3 q_4} = \frac{1}{4}C_F\, \hat{P}^{\mu\nu\,(\text{ab})}_{\bar{q}^\prime_1 q^\prime_2 \bar q_3 q_4}+ \frac{1}{4}C_A\, \hat{P}^{\mu\nu\,(\text{nab})}_{\bar{q}^\prime_1 q^\prime_2 \bar q_3 q_4}\,,
\end{equation}
 where the indices carried by the parton label refer to the indices of the momenta and the momentum fractions of the partons, and $C_F$ and $C_A$ denote the quadratic Casimirs of the fundamental and adjoint representations of SU$(N)$,
 \begin{equation}
     C_F = \frac{N^2-1}{2N}\,,\qquad C_A = N\,.
 \end{equation}

The splitting process $g \to \bar q q \bar q q$, in which the final state quarks have the same flavour, $q'=q$, includes diagrams where the anti-quarks 1 and 3 are exchanged (or equivalently, where the quarks 2 and 4 are exchanged). This naturally leads to the following representation of the splitting amplitude,
\begin{equation}
\hat{P}^{\mu\nu}_{\bar q_1 q_2 \bar q_3 q_4} =  \left[ \hat{P}^{\mu\nu}_{\bar q'_1 q'_2 \bar q_3 q_4} 
+ \left( 1 \leftrightarrow 3 \right) \right] 
+ \hat{P}^{\mu\nu\,(\text{id})}_{\bar q_1 q_2 \bar q_3 q_4}\,.
\label{eq:Pqqqq}
\end{equation}
The term in square brackets contains the splitting amplitude in eq.~\eqref{eq:Pq_qb'q'qbq} in the case of different flavours and the exchange contributions obtained by permuting the external quarks. The last term in eq.~\eqref{eq:Pqqqq} is new and captures interference contributions from identical quarks. It is again convenient to display the result in terms of colour factors,
\begin{align}
\hat{P}^{\mu\nu\,(\text{id})}_{\bar q_1 q_2 \bar q_3 q_4} &=
\frac{1}{2}C_F\,(C_A-2C_F)\,\hat{P}^{\mu\nu\,(\text{id})_1}_{\bar q_1 q_2 \bar q_3 q_4} +
\frac{1}{2}C_A\,(C_A-2C_F)\,\hat{P}^{\mu\nu\,(\text{id})_2}_{\bar q_1 q_2 \bar q_3 q_4}\,.
\label{eq:P_qqqq_id}
\end{align}
Since $C_A-2C_F=\frac{1}{N}$, the interference contributions are colour suppressed.
%
%This form is particularly useful when studying the soft $q \bar q$ emission within the four-parton collinear set, or when verifying the N$^3$LO supersymmetric Ward identity.

\iffalse
%
\begin{align}
\hat{P}^{\mu\nu\,(\text{id})}_{\bar q_1 q_2 \bar q_3 q_4} &=
\frac{1}{4}C_F\,\hat{P}^{\mu\nu\,(\text{id})(\text{ab})_1}_{\bar q_1 q_2 \bar q_3 q_4} +
\frac{1}{4}C_A\,\hat{P}^{\mu\nu\,(\text{id})(\text{nab})_1}_{\bar q_1 q_2 \bar q_3 q_4} +
\frac{1}{2}C_F^2\,\hat{P}^{\mu\nu\,(\text{id})(\text{ab})_2}_{\bar q_1 q_2 \bar q_3 q_4} \\
&+\frac{1}{2}C_A^2\,\hat{P}^{\mu\nu\,(\text{id})(\text{nab})_2}_{\bar q_1 q_2 \bar q_3 q_4} +
\frac{1}{2}C_A C_F\,\hat{P}^{\mu\nu\,(\text{id})(\text{id})}_{\bar q_1 q_2 \bar q_3 q_4}\,.
\label{eq:P_qqqq_id}
\end{align}
%
\fi

%%%%%%%%%%%%%%% g -> qb g g q %%%%%%%%%%%%%%%%%%%%%
\begin{figure}[!h]
\centering
\begin{align}&
\begin{tikzpicture}[baseline=(a)]
\begin{feynman}[inline=(a)]
\node[blob, scale=1] (blob1);
\vertex [left=0.5cm of blob1] (bracket){\(\Big\{\)};
\vertex [right=0.3cm of blob1] (a);
\vertex [above right = 0.3cm of a, draw=none] (null1){\(\)};
\vertex [right=0.5cm of a] (b);
\vertex [above right= 1.9cm of b] (g2) {\(g_2\)};
\vertex [right=0.6cm of b] (c);
\vertex [above right= 1.9cm of c] (g3) {\(g_3\)};
\vertex [right=0.7cm of c] (d);
\vertex [above right= 1.9cm of d] (q1) {\(\bar{q}_1\)};
\vertex [below right=1.9cm of d] (q4) {\(q_4\)};
\vertex [right=1.3cm of d, draw=none] (plus){\(+\)};
\diagram*{
(blob1)-- (a) -- [gluon] (d),
(b) -- [gluon] (g2),
(c) -- [gluon] (g3),
(d) -- [anti fermion] (q1),
(d) -- [fermion] (q4),
};
\end{feynman}
\end{tikzpicture} 
\begin{tikzpicture}[baseline=(a)]
\begin{feynman}[inline=(a)]
\node[blob, scale=1] (blob1);
\vertex [right=0.3cm of blob1] (a);
\vertex [above right = 0.3cm of a, draw=none] (null1){\(\)};
\vertex [right=0.5cm of a] (b);
\vertex [above right= 1.9cm of b] (g2) {\(g_2\)};
\vertex [right=1cm of b] (c);
\vertex [above right= 1.9cm of c] (q1) {\(\bar{q}_1\)};
\vertex [below right=1cm of c] (q41);
\vertex [above right=1.5cm of q41] (g3){\(g_3\)};
\vertex [below right=0.9cm of q41] (q42) {\(q_4\)};
\vertex [right=2cm of c, draw=none] (plus){\(+\)};
\diagram*{
(blob1)-- (a) -- [gluon] (c),
(b) -- [gluon] (g2),
(c) -- [anti fermion] (q1),
(c) -- [fermion] (q41) -- [fermion] (q42),
(q41) -- [gluon] (g3)
};
\end{feynman}
\end{tikzpicture} 
\begin{tikzpicture}[baseline=(a)]
\begin{feynman}[inline=(a)]
\node[blob, scale=1] (blob1);
\vertex [right=0.3cm of blob1] (a);
\vertex [above right = 0.3cm of a, draw=none] (null1){\(\)};
\vertex [right=0.5cm of a] (b);
\vertex [above right= 1.9cm of b] (g2) {\(g_2\)};
\vertex [right=1cm of b] (c);
\vertex [above right=1cm of c] (q11);
\vertex [above right= 0.9cm of q11] (q12) {\(\bar{q}_1\)};
\vertex [below right=1.5cm of q11] (g3){\(g_3\)};
\vertex [below right=1.9cm of c] (q4) {\(q_4\)};
\vertex [right=2cm of c, draw=none] (plus){\(+\quad (2\leftrightarrow3) \Big\}\)};
\diagram*{
(blob1)-- (a) -- [gluon] (c),
(b) -- [gluon] (g2),
(c) -- [anti fermion] (q11) -- [anti fermion] (q12),
(c) -- [fermion] (q4),
(q11) -- [gluon] (g3)
};
\end{feynman}
\end{tikzpicture} \nonumber\\ &
\begin{tikzpicture}[baseline=(a)]
\begin{feynman}[inline=(a)]
\node[blob, scale=1] (blob1);
\vertex [left=0.5cm of blob1] (pl){\(+\)};
\vertex [right=0.3cm of blob1] (a);
\vertex [above right = 0.3cm of a, draw=none] (null1){\(\)};
\vertex [right=1.1cm of a] (b);
\vertex [above right=1cm of b] (c);
\vertex [above right=0.9 cm of c] (q1){\(\bar{q}_1\)};
\vertex [below right=1.5cm of c] (g3){\(g_3\)};
\vertex [below right = 0.5cm of c, draw=none] (null2);
\vertex [above right = 1cm of null2] (g2){\(g_2\)};
\vertex [below right=1.9cm of b] (q4) {\(q_4\)};
\vertex [right=2cm of b, draw=none] (plus){\(+\)};
\diagram*{
(blob1) -- (a) -- [gluon] (b) -- [anti fermion] (c),
(c) -- [anti fermion] (q1),
(c) -- [gluon] (g3),
(null2) -- [gluon] (g2),
(b) -- [fermion] (q4)
};
\end{feynman}
\end{tikzpicture} 
\begin{tikzpicture}[baseline=(a)]
\begin{feynman}[inline=(a)]
\node[blob, scale=1] (blob1);
\vertex [right=0.3cm of blob1] (a);
\vertex [above right = 0.3cm of a, draw=none] (null1){\(\)};
\vertex [right=1.1cm of a] (b);
\vertex [above right=1.9 cm of b] (q1){\(\bar{q}_1\)};
\vertex [below right=1cm of b] (c);
\vertex [above right=1.5cm of c] (g2){\(g_2\)};
\vertex [above right = 0.5cm of c,draw=none] (null2);
\vertex [below right = 1cm of null2] (g3){\(g_3\)};
\vertex [below right=0.9cm of c] (q4) {\(q_4\)};
\vertex [right=2cm of b, draw=none] (plus){\(+\)};
\diagram*{
(blob1) -- (a) -- [gluon] (b) -- [anti fermion] (q1),
(b) -- [fermion] (c),
(c) -- [fermion] (q4),
(c) -- [gluon] (g2),
(null2) -- [gluon] (g3)
};
\end{feynman}
\end{tikzpicture} 
\begin{tikzpicture}[baseline=(a)]
\begin{feynman}[inline=(a)]
\node[blob, scale=1] (blob1);
\vertex [right=0.3cm of blob1] (a);
\vertex [above right = 0.3cm of a, draw=none] (null1){\(\)};
\vertex [right=0.5cm of a] (b);
\vertex [above right= 1.9cm of b] (g2) {\(g_2\)};
\vertex [right=1cm of b] (c);
\vertex [above right=1cm of b,draw=none] (null2);
\vertex [right= 1cm of null2] (g3) {\(g_3\)};
\vertex [right=1cm of c] (d);
\vertex [above right= 1.9cm of d] (q1) {\(\bar{q}_1\)};
\vertex [below right=1.9cm of d] (q4) {\(q_4\)};
\diagram*{
(blob1)-- (a) -- [gluon] (d),
(b) -- [gluon] (g2),
(null2) -- [gluon] (g3),
(d) -- [anti fermion] (q1),
(d) -- [fermion] (q4)
};
\end{feynman}
\end{tikzpicture} \nonumber\\&
\begin{tikzpicture}[baseline=(a)]
\begin{feynman}[inline=(a)]
\node[blob, scale=1] (blob1);
\vertex [right=0.3cm of blob1] (a);
\vertex [above right = 0.3cm of a, draw=none] (null1){\(\)};
\vertex [left=0.5cm of blob1] (bracket){\(\Big\{\)};
\vertex [left=1cm of blob1] (pl){\(+\)};
\vertex [right=1.1cm of a] (b);
\vertex [above right=1 cm of b] (q11);
\vertex [above right=0.9 cm of q11] (q12){\(\bar{q}_1\)};
\vertex [below right = 0.7cm of q11] (g2){\(\)};
\vertex [above right = 0.4cm of b] (n1);
\vertex [right = 0.9cm of n1] (n2) {\(g_2\)};
\vertex [below right=1 cm of b] (q41);
\vertex [below right=0.9 cm of q41] (q42){\(q_4\)};
\vertex [above right = 0.7cm of q41] (g3){\(\)};
\vertex [below right = 0.4cm of b] (m1);
\vertex [right = 0.9cm of m1] (m2) {\(g_3\)};
\vertex [right=1.7cm of b, draw=none] (plus){\(+\)};
\diagram*{
(blob1) -- (a) -- [gluon] (b) -- [anti fermion] (q11)-- [anti fermion] (q12),
(b) -- [fermion] (q41) -- [fermion] (q42) ,
(q11) -- [gluon] (g2),
(q41) -- [gluon] (g3),
};
\end{feynman}
\end{tikzpicture} 
\begin{tikzpicture}[baseline=(a)]
\begin{feynman}[inline=(a)]
\node[blob, scale=1] (blob1);
\vertex [right=0.3cm of blob1] (a);
\vertex [above right = 0.3cm of a, draw=none] (null1){\(\)};
\vertex [right=1.1cm of a] (b);
\vertex [above right=0.7 cm of b] (q11);
\vertex [above right=0.7 cm of q11] (q12);
\vertex [above right=0.5 cm of q12] (q13){\(\bar{q}_1\)};
\vertex [below right = 0.8cm of q11] (g3){\(g_3\)};
\vertex [below right = 0.8cm of q12] (g2){\(g_2\)};
\vertex [below right=1.9 cm of b] (q4){\(q_4\)};
\vertex [right=2cm of b, draw=none] (plus){\(+\)};
\diagram*{
(blob1) -- (a) -- [gluon] (b) -- [anti fermion] (q11)-- [anti fermion] (q12) -- [anti fermion] (q13),
(b) -- [fermion] (q4),
(q11) -- [gluon] (g3),
(q12) -- [gluon] (g2),
};
\end{feynman}
\end{tikzpicture} 
\begin{tikzpicture}[baseline=(a)]
\begin{feynman}[inline=(a)]
\node[blob, scale=1] (blob1);
\vertex [right=0.3cm of blob1] (a);
\vertex [above right = 0.3cm of a, draw=none] (null1){\(\)};
\vertex [right=1.1cm of a] (b);
\vertex [above right=1.9 cm of b] (q1){\(\bar{q}_1\)};
\vertex [below right=0.7 cm of b] (q41);
\vertex [below right=0.7 cm of q41] (q42);
\vertex [below right=0.5 cm of q42] (q43){\(q_4\)};
\vertex [above right = 0.8cm of q41] (g2){\(g_2\)};
\vertex [above right = 0.8cm of q42] (g3){\(g_3\)};
\vertex [right=2cm of b, draw=none] (plus){\(+\quad (2\leftrightarrow3) \Big\}\)};
\diagram*{
(blob1) -- (a) -- [gluon] (b) -- [fermion] (q41)-- [fermion] (q42) -- [fermion] (q43),
(b) -- [anti fermion] (q1),
(q41) -- [gluon] (g2),
(q42) -- [gluon] (g3),
};
\end{feynman}
\end{tikzpicture} \nonumber\\& 
\begin{tikzpicture}[baseline=(a)]
\begin{feynman}[inline=(a)]
\node[blob, scale=1] (blob1);
\vertex [right=0.3cm of blob1] (a);
\node[blob, scale=1] (blob1);
\vertex [left=0.5cm of blob1] (pl){\(+\)};
\vertex [above right = 0.3cm of a, draw=none] (null1){\(\)};
\vertex [right=0.5cm of a] (b);
\vertex [above right= 1.9cm of b] (g2) {\(g_2\)};
\vertex [below right= 1.9cm of b] (g3) {\(g_3\)};
\vertex [right=1cm of b] (d);
\vertex [above right= 1.9cm of d] (q1) {\(\bar{q}_1\)};
\vertex [below right=1.9cm of d] (q4) {\(q_4\)};
\diagram*{
(blob1)-- (a) -- [gluon] (d),
(b) -- [gluon] (g2),
(b) -- [gluon] (g3),
(d) -- [anti fermion] (q1),
(d) -- [fermion] (q4)
};
\end{feynman}
\end{tikzpicture} \nonumber
\end{align}
\caption{The diagrams contributing to the collinear decay $g \to \bar{q}_1 g_2 g_3 q_4$.}
\label{fig:qggq}
\end{figure}
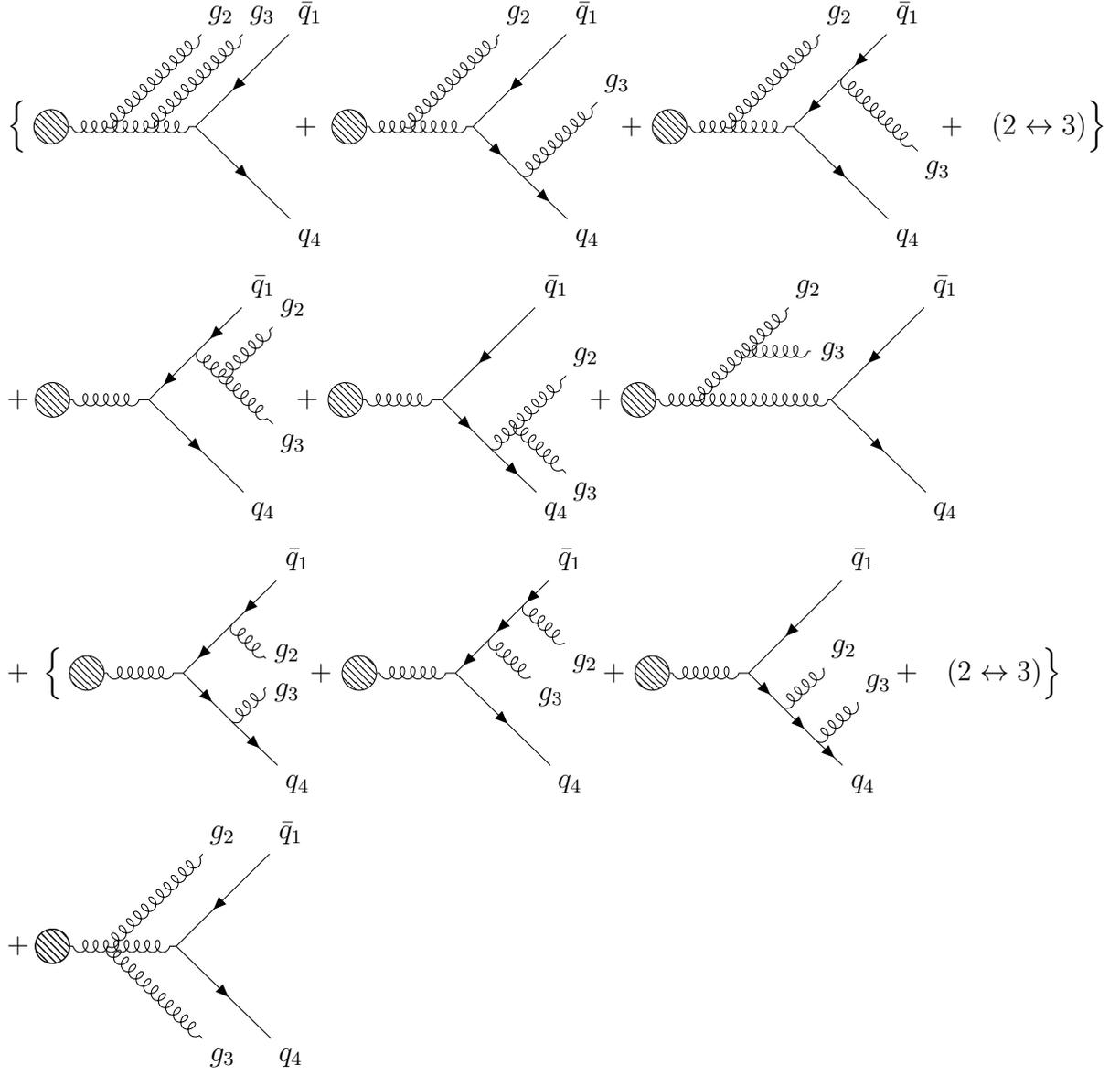

Next, let us discuss the splitting amplitude $g \to \bar q q g g$. The Feynman diagrams contributing to the off-shell current $\textrm{J}^\mu_{\bar q q g g}$ are shown in fig.~\ref{fig:qggq}. As usual, we can decompose the splitting amplitude into contributions from different colour factors as follows,
\begin{align}
    \hat P^{\mu\nu}_{\bar q_1 g_2 g_3 q_4} =
    \frac{1}{2}C_F^2\, \hat P^{\mu\nu\,(\text{ab})}_{\bar q_1 g_2 g_3 q_4}
    +\frac{1}{2}C_A^2\, \hat P^{\mu\nu\,(\text{nab})_1}_{\bar q_1 g_2 g_3 q_4}
    +\frac{1}{2}C_A\,C_F\, \hat P^{\mu\nu\,(\text{nab})_2}_{\bar q_1 g_2 g_3 q_4}\,.
\label{eq:g_qggq}
\end{align}
It is possible to express each colour coefficient in eq.~\eqref{eq:g_qggq} in a reduced form by exploiting the symmetry under the exchange of the two external partons,
\begin{align}
    \hat P^{\mu\nu\,(X)}_{\bar q_1 g_2 g_3 q_4} =
    \left(
    \hat P^{\mu\nu\,(X)\,\text{ symm.}}_{\bar q_1 g_2 g_3 q_4}
    + (1\leftrightarrow 4)
    \right)
    + (2\leftrightarrow 3)\,,
\label{eq:g_qggq_symm}
\end{align}
where $(X) \in \{(\text{ab}),(\text{nab})_1,(\text{nab})_2\}$.

\iffalse
The decomposition of the symmetrised expression $\hat P^{\mu\nu\,\text{ symm.}}_{\bar q_1 g_2 g_3 q_4}$ into different colour factors follows the same pattern as that of eq.~\eqref{eq:g_qggq}, \rayan{added, referenced on website}
%
\begin{align}
       \hat P^{\mu\nu\,\text{ symm.}}_{\bar q_1 g_2 g_3 q_4} &=
    \frac{1}{2}C_F^2\, \hat P^{\mu\nu\,(\text{ab})\,\text{ symm.}}_{\bar q_1 g_2 g_3 q_4}
    +\frac{1}{2}C_A^2\, \hat P^{\mu\nu\,(\text{nab})_1\,\text{ symm.}}_{\bar q_1 g_2 g_3 q_4} \nonumber \\
    &+\frac{1}{2}C_A\,C_F\, \hat P^{\mu\nu\,(\text{nab})_2\,\text{ symm.}}_{\bar q_1 g_2 g_3 q_4}\,.
\end{align}
%
\fi

\begin{figure}[!h]
\centering
\begin{equation}
\begin{split}
&
\scalebox{0.9}{
\begin{tikzpicture}[baseline=(a)]
\begin{feynman}[inline=(a)]
\node[blob, scale=1] (blob1);
\vertex [left=0.5cm of blob1] (bracket){\(\Big\{\)};
\vertex [right=0.3cm of blob1] (a);
\vertex [below = 0.1cm of a, draw=none] (null1);
\vertex [right=1.2cm of a] (d1);
\vertex [above right = 0.8cm of d1] (null2);
\vertex [above right= 2.2cm of d1] (g1) {\(g_1\)};
\vertex [above right = 1.5cm of d1] (null3);
\vertex [below right= 1.4cm of null2] (g3) {\(g_3\)};
\vertex [below right= 0.8cm of null3] (g2) {\(g_2\)};
\vertex [below right=2.2cm of d1] (g4){\(g_4\)};
\vertex [right=2.2cm of d1, draw=none] (plus){\(+\,11 \text{ permutations}\Big\}\)};
\diagram*{
(blob1)-- (a) -- [gluon]  (d1) -- [gluon] (g1), %, momentum'={\small\(p_{1234}\)}
(null2) -- [gluon] (g3),
(null3) -- [gluon] (g2),
(d1) -- [gluon] (g4),
};
\end{feynman}
\end{tikzpicture}}
%%%%%%%%
\scalebox{0.9}{
\begin{tikzpicture}[baseline=(a)]
\begin{feynman}[inline=(a)]
\node[blob, scale=1] (blob1);
\vertex [left=0.8cm of blob1] (br){\(+\quad\Big\{\)};
\vertex [right=0.3cm of blob1] (a);
\vertex [below = 0.1cm of a, draw=none] (null1);
\vertex [right=1.2cm of a] (d1);
\vertex [above right = 1.5cm of d1] (null2);
\vertex [below right = 1.5cm of d1] (null3);
\vertex [above right= 2.3cm of d1] (g1) {\(g_1\)};
\vertex [below right= 0.8cm of null2] (g2) {\(g_2\)};
\vertex [above right= 0.8cm of null3] (g3) {\(g_3\)};
\vertex [below right=2.3cm of d1] (g4){\(g_4\)};
\vertex [right=2.2cm of d1, draw=none] (plus){\(+\,(1\leftrightarrow 3) + (2\leftrightarrow 3)\Big\}\)};
\diagram*{
(blob1)-- (a) -- [gluon]  (d1) -- [gluon] (g1), %, momentum'={\small\(p_{1234}\)}
(null2) -- [gluon] (g2),
(null3) -- [gluon] (g3),
(d1) -- [gluon] (g4)
};
\end{feynman}
\end{tikzpicture}} \\
%%%%%%%%
&
\scalebox{0.9}{
\begin{tikzpicture}[baseline=(a)]
\begin{feynman}[inline=(a)]
\node[blob, scale=1] (blob1);
\vertex [right=0.3cm of blob1] (a);
\vertex [left=1.2cm of blob1] (pl){\(+\)};
\vertex [left=0.7cm of blob1,yshift=-0.2cm] (sum2){\(\mathlarger{\sum}\limits_{\sigma \in S_2 \text{ in } S_4}\)};
\vertex [below = 0.1cm of a, draw=none] (null1);
\vertex [right=1.2cm of a] (d1);
\vertex [above right = 1.5cm of d1] (null2);
\vertex [above right = 0.8cm of d1] (null3);
\vertex [right = 1.05cm of null3] (null4){\(g_{\sigma(2)}\)};
\vertex [above right= 2.2cm of d1] (g1) {\(g_{\sigma(1)}\)};
\vertex [right= 1.6cm of d1] (g3) {\(g_{\sigma(3)}\)};
\vertex [below right= 0.8cm of null2] (g2);
\vertex [below right=2.2cm of d1] (g4){\(g_{\sigma(4)}\)};
\diagram*{
(blob1)-- (a) -- [gluon]  (d1) -- [gluon] (g1), %, momentum'={\small\(p_{1234}\)}
(null2) -- [gluon] (g2),
(d1) -- [gluon] (g3),
(d1) -- [gluon] (g4)
};
\end{feynman}
\end{tikzpicture}}
%%%%%%%
\scalebox{0.9}{
\begin{tikzpicture}[baseline=(a)]
\begin{feynman}[inline=(a)]
\node[blob, scale=1] (blob1);
\vertex [left=0.8cm of blob1] (br){\(+\quad\Big\{\)};
\vertex [right=0.3cm of blob1] (a);
\vertex [below = 0.1cm of a, draw=none] (null1);
\vertex [right=1.2cm of a] (d1);
\vertex [above right = 1.2cm of d1] (null2);
\vertex [above right= 2.3cm of d1] (g1) {\(g_1\)};
\vertex [right= 0.8cm of null2] (g2) {\(g_2\)};
\vertex [below right= 1cm of null2] (g3) {\(g_3\)};
\vertex [below right=2.3cm of d1] (g4){\(g_4\)};
\vertex [right=2.2cm of d1, draw=none] (plus){\(+\,(1\leftrightarrow 4) + (2\leftrightarrow 4) + (3\leftrightarrow 4)\Big\}\)};
\diagram*{
(blob1)-- (a) -- [gluon]  (d1) -- [gluon] (g1), %, momentum'={\small\(p_{1234}\)}
(null2) -- [gluon] (g2),
(null2) -- [gluon] (g3),
(d1) -- [gluon] (g4)
};
\end{feynman}
\end{tikzpicture}}
\end{split} \nonumber
\end{equation}
\label{fig:g_gggg}
\caption{The diagrams contributing to the splitting process $g\to g_1g_2g_3g_4$. In the third diagram we sum over the 6 possible pairings of the partons in the three-gluon vertex.}
%The group $S_2$ in $S_4$ includes the 6 possible pairings of 2 elements in a list of 4 elements.
\end{figure}
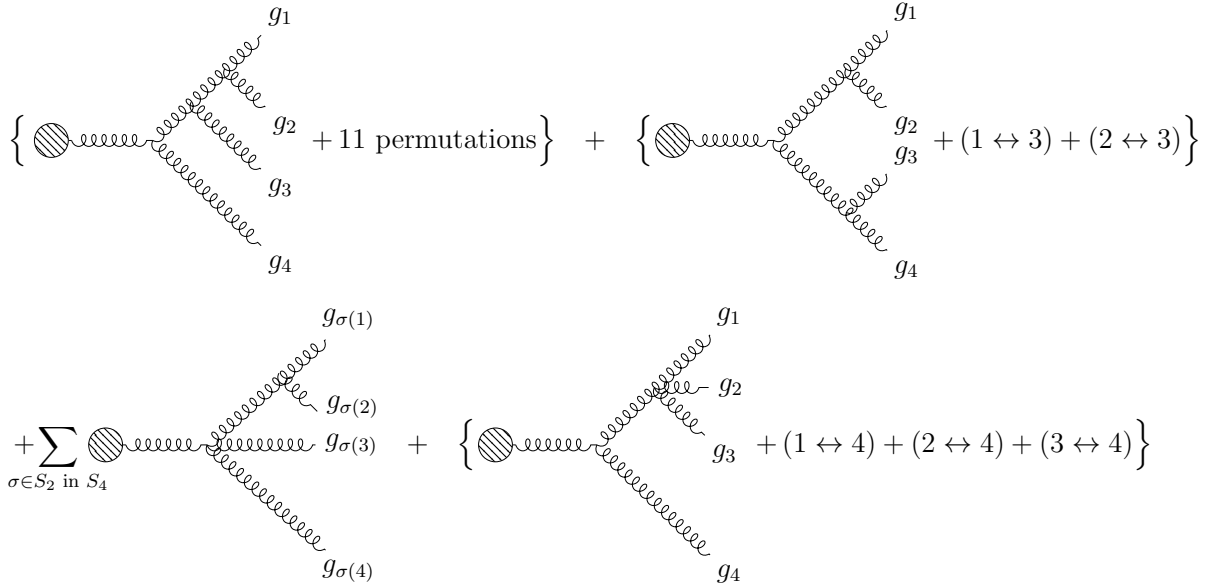

Finally, let us discuss the pure gluon splitting process $g \to g g g g$, which poses a challenge due to the large degree of Bose symmetry under the exchange of the external gluons. The diagrams contributing to the decay are shown in fig.~\ref{fig:g_gggg}. An important step in the computation of the splitting amplitude was to take into account symmetries between different permutatons of the four external gluons in order to minimise the number of terms. We can write $\hat{P}
^{\mu\nu}_{gggg}$ in a symmetrised form as
\begin{equation}
\hat{P}^{\mu\nu}_{g_1g_2g_3g_4} =
\hat{P}^{\mu\nu\,\text{ symm.}}_{g_1g_2g_3g_4}
+ (11 \text{ permutations of } g_1g_2g_3g_4 )\,.
\label{eq:g_gggg_symm}
\end{equation}
The above permutations do not include orderings of the external gluons which leave the first diagram in fig.~\ref{fig:g_gggg} invariant.
\iffalse
Fixing the reference ordering to be $(1,2,3,4)$, which label the external gluons of the diagrams in fig.~\ref{fig:g_gggg} top to bottom, the permutations read
%
\begin{equation}
\begin{split}
\text{11 permutations of } (1,2,3,4) = 
\begin{cases}
&(3, 2, 1, 4), \\
&(4, 2, 3, 1),\\
&(1, 3, 2, 4), \\
&(1, 4, 3, 2), \\
&(1, 2, 4, 3), \\
&(1, 3, 4, 2), \\
&(1, 4, 2, 3), \\
&(3, 2, 4, 1), \\
&(4, 2, 1, 3), \\
&(3, 4, 1, 2), \\
&(3, 4, 2, 1).
% %%%%%%%
%&(1 \leftrightarrow 3), \\
%&(1 \leftrightarrow 4),\\
%&(2 \leftrightarrow 3), \\
%&(2 \leftrightarrow 4), \\
%&(3 \leftrightarrow 4), \\
%&(2 \leftrightarrow 3) +  (2 \leftrightarrow 4), \\
%&(2 \leftrightarrow 4) +  (2 \leftrightarrow 3), \\
%&(1 \leftrightarrow 3) +  (1 \leftrightarrow 4), \\
%&(1 \leftrightarrow 4) +  (1 \leftrightarrow 3), \\
%&(1 \leftrightarrow 3) +  (2 \leftrightarrow 4), \\
%&(1 \leftrightarrow 3) +  (2 \leftrightarrow 4) + (1 \leftrightarrow 2)\,,
%%%%%%%%%
\end{cases}
\end{split}
\end{equation}
%
\fi

\section{Nested collinear limits}
\label{sec:nested_collinear}
%\subsection{Strongly-ordered vs. iterated collinear limits}

In this section we analyse the collinear limit of the splitting amplitudes themselves, i.e., we study their behaviour in the limit where a subset of collinear partons is more collinear than the others. 
To be concrete, let us consider a collection of $m$ partons with flavour indices $\{f_1,\ldots, f_{m'},\ldots,f_m\}$ and momenta $\{p_1,\ldots,p_{m'},\ldots,p_m\}$, with $m'<m$. We always think of these partons as being part of an on-shell $n$-point amplitude $\mathcal M_{f_1 \ldots f_n}$ involving $(n-m)$ additional coloured partons. Our goal is to study the behaviour of the amplitude in the limit where $\{p_1,\ldots,p_{m'}\}$ become collinear to some lightlike direction $\pN{}$, and $\{\pN{},p_{m'+1},\ldots, p_m\}$ are collinear to another lightlike direction $\pM{}$. Depending on the order in which the different collinear limits are taken, there are two different scenarios of how such a kinematic configuration can be reached, referred to as \emph{iterated} and \emph{strongly-ordered} collinear limits in ref.~\cite{DelDuca:2019ggv}. The (splitting) amplitudes factorise in the same way in each of the limits, and the factorisation involves the same universal quantities in both cases~\cite{DelDuca:2019ggv}. We therefore focus here only on the strongly-ordered limit from now on. 

We start by giving a precise definition of the strongly-ordered collinear limit.
We perform separate light-cone decompositions in each of the $m$- and $m'$-parton sets. For the $m$-parton set, we will use the notations and conventions of eq.~\eqref{eq:collinear_momenta}.
For the $m'$-parton subset we write
\begin{equation}
p^\mu_i = y_i \pN{\mu} + \Kb{i}^\mu - \frac{\Kb{i}^2}{2 y_i}\frac{n'^\mu}{n' \cdot \pN{}}\,, 
\quad i= 1,\ldots,m'\,,
\label{eq:Sudakov1.2}
\end{equation}
with $n'^2 = \pN{2} = \pN{}\cdot\Kb{i} = n'\cdot \Kb{i}=0$. 
The momenta $\pM{}$ and $\pN{}$ indicate the directions to which the partons in each set become collinear. We stress that at this point the lightcone directions $\pN{}$ and $n'$ in eq.~\eqref{eq:Sudakov1.2} are not related to the quantities $\pM{}$ and $n$ in eq.~\eqref{eq:collinear_momenta}.
%and $\pN{}$ in eq.~\eqref{eq:Sudakov1.2} specify different light-cone directions, $\pM{}\cdot \K{i} =0= \pN{}\cdot \Kb{i}$, while the auxiliary light-like vectors $n^\mu$ and $n'^\mu$ specify how the collinear directions are approached, $n \cdot \K{i}  = 0= n' \cdot \Kb{i} $. 
However, without loss of generality, we may choose $n'=n$, and we work in the axial gauge where the reference vectors of all external and internal gluons are $n$. For more details about the parametrisation of the strongly-ordered collinear limit, we refer to ref.~\cite{DelDuca:2019ggv}.
%Indeed, if $\pM{} = E (1,\vec{u})$ and $\pN{}=E'(1,\vec{u'})$ are given, we can choose any lightlike vectors $n$ and $n'$ such that $n\cdot \pM{}\neq0$ and $n'\cdot \pN{}\neq0$. For example, we may choose $n=(1,\vec v)$ and $n'=(1,\vec {v'})$, where $\vec{v}$ and $\vec{v'}$ are unit vectors. The choice of these unit vectors is arbitrary, as long as $\vec{u}\cdot\vec{v}\neq1$ and $\vec{u'}\cdot\vec{v'}\neq1$. It is then easy to see that we can always assume without loss of generality $\vec{v}=\vec{v'}$, i.e., $n=n'$. Let us also mention that, just like in the case of the ordinary collinear limit in Sec.~\ref{sec:setup},
%In App.~\ref{app:nested} we show that we may choose $n'^\mu = n^\mu$. 
%
With this setup, the strongly-ordered collinear limit is defined in analogy with the ordinary collinear limit in Sec.~\ref{sec:setup}: the vectors $\K{i}^\mu$ and $\Kb{i}^\mu$ parametrise the transverse distance to the planes spanned by $(\pM{},n)$ and $(\pN{},n)$, respectively. The strongly-ordered collinear limit where the $m'$-parton subset is more collinear than the $m$-parton set is defined as the limit where both $\K{i}^\mu$ and $\Kb{i}^\mu$ approach zero, but the $\Kb{i}^\mu$ tend to zero faster than the $\K{i}^\mu$. We can implement the operation of taking this limit by a uniform rescaling of the transverse momenta in each collinear set by a different parameter,
\begin{align}
   &\K{i} \to \lambda\, \K{i}, \quad \Kb{i} \to \lambda'\, \Kb{i},
   \label{eq:so_rescaling}
\end{align}
and keeping the dominant singular terms of order $1/\lambda'^{\,2(m' -1)}\lambda^{2(m-m')}$ in the limit $\lambda, \lambda' \to 0$ with $ \lambda \gg \lambda'$. 

The leading behaviour of an amplitude in the strongly-ordered collinear limit is described by a factorisation formula very similar to eq.~\eqref{eq:factorisation}~\cite{DelDuca:2019ggv},
{\begin{align} \label{eq:H_def}
\mathscr{C}_{(1\ldots m')\ldots m} & \mathscr{C}_{1\ldots m'} 
|\mathcal{M}_{f_1 \ldots f_n}(p_1,\ldots,p_n)|^2 = \left(\frac{2g_s^2\mu^{2\epsilon}}{s_{1\ldots m'}}\right)^{m'-1} \left(\frac{2g_s^2\mu^{2\epsilon}}{s_{[1\ldots m']\ldots m}}\right)^{m-m'} \nonumber\\
&\times 
\hat{P}^{hh'}_{f_1\ldots f_{m'}}\,\hat{H}^{hh';ss'}_{f_{(1\ldots m')}f_{m'+1}\ldots f_m}\,\mathcal{T}_{ff_{m+1}\ldots f_n}^{ss'}(\pM{},p_{m'+1},\ldots,p_n)\,,
\end{align}}
where 
\begin{align}
    s_{[1\ldots m']\ldots m} = (\pN{}+p_{m'+1}+\ldots +p_m)^2\,.
    \label{eq:s_square}
\end{align}
The functions $\hat{P}^{hh'}_{f_1\ldots f_{m'}}$ and $\mathcal{T}_{ff_{m+1}\ldots f_n}^{ss'}$ are the splitting amplitude and the helicity tensor introduced in Sec.~\ref{sec:setup}. The \emph{splitting tensor} $\hat{H}^{hh';ss'}_{f_{(1\ldots m')}f_{m'+1}\ldots f_m}$ is new. It is obtained by squaring the amplitude-level splitting amplitude without summing over the helicities of one of the partons in the collinear set (cf.~eq.~\eqref{eq:P_def}),
\beq\bsp\label{eq:H_Sp_Def}
&\left(\frac{2g_s^2\mu^{2\epsilon}}{s_{[1\ldots m']\ldots m}}\right)^{m-m'}\,\hat{H}^{hh';ss'}_{f_{(1\ldots m')}f_{m'+1}\ldots f_m} \\
&\qquad = \frac1{\mathcal{C}_{f}}
\sum_{\substack{(s_{m'+1},\ldots,s_m)\\(c,c_{m'+1},\ldots,c_m)}}  {\bf Sp}_{f_{(1\ldots m')}f_{m'+1}\ldots f_m}^{c,c_{m'+1}\ldots c_m;s,h,s_{m'+1}\ldots s_m}\,\left[{\bf Sp}_{f_{(1\ldots m')}f_{m'+1}\ldots f_m}^{c,c_{m'+1}\ldots c_m;s',h',s_{m'+1}\ldots s_m}\right]^{\ast}\,,
\esp\eeq
where $\mathcal{C}_{f}$ is defined after eq.~(\ref{eq:P_def}).
Just like in Sec.~\ref{sec:setup} we suppress the dependence of all splitting amplitudes and tensors on their arguments. The factorisation of the squared amplitude in the strongly-ordered limit can be cast in the form of a factorisation of the splitting amplitude itself,
\beq\label{eq:P_coll}
\mathscr{C}_{1\ldots m'} \hat{P}^{ss'}_{f_1\ldots f_{m}}= \left(\frac{s_{[1\ldots m']\ldots m}}{s_{1\ldots m'}}\right)^{m'-1} \,
\hat{P}^{hh'}_{f_1\ldots f_{m'}}\,\hat{H}^{hh';ss'}_{f_{(1\ldots m')}f_{m'+1}\ldots f_m}\,.
\eeq
By comparing eqs.~\eqref{eq:factorisation} and~\eqref{eq:H_def}, it is easy to see that upon summing over the helicities $(h,h')$ the splitting tensor reduces to an ordinary splitting amplitude,
\beq\label{eq:H_to_P_helicities}
\delta^{hh'}\,\hat{H}^{hh';ss'}_{f_{(1\ldots m')}f_{m'+1}\ldots f_m} = \hat{P}^{ss'}_{f_{(1\ldots m')}f_{m'+1}\ldots f_m}\,.
\eeq

In the following we refer to the partons with spin indices $(s,s')$ and $(h,h')$ as the \emph{parent} and \emph{sub-parent}, respectively.
Depending on the flavour of the parent and the sub-parent, the structure of the splitting tensor can be further simplified. The case where the parent is a quark was considered in ref.~\cite{DelDuca:2019ggv}. Here we only consider the case where the parent is a gluon. Using a similar argument as for the splitting amplitude in Sect.~\ref{sec:collinear}, we can trade helicity indices $(s,s')$ for Lorentz indices $(\mu,\nu)$ and write eq.~\eqref{eq:H_def} in the equivalent form,
\begin{align} 
\label{eq:P_coll_g}
\mathscr{C}_{(1\ldots m')\ldots m} & \mathscr{C}_{1\ldots m'} 
|\mathcal{M}_{f_1 \ldots f_n}(p_1,\ldots,p_n)|^2 = \left(\frac{2g_s^2\mu^{2\epsilon}}{s_{1\ldots m'}}\right)^{m'-1} \left(\frac{2g_s^2\mu^{2\epsilon}}{s_{[1\ldots m']\ldots m}}\right)^{m-m'} \nonumber\\
&\times 
\hat{P}^{hh'}_{f_1\ldots f_{m'}}\,\hat{H}^{hh';\mu\nu}_{f_{(1\ldots m')}f_{m'+1}\ldots f_m}\,\mathcal{T}_{ff_{m+1}\ldots f_n,\mu\nu}(\pM{},p_{m'+1},\ldots,p_n)\,,
\end{align}
%
%%%%%%%%%%%%%%%
If the sub-parent is a quark, helicity must be conserved, and the splitting tensor is diagonal in the spin indices $(h,h')$ of the sub-parent. 
Equation~\eqref{eq:H_to_P_helicities} then implies
\beq\label{eq:Hg_qf}
\mathscr{C}_{1\ldots m'} \hat{P}^{\mu\nu}_{f_1\ldots f_{m}}= \left(\frac{s_{[1\ldots m']\ldots m}}{s_{1\ldots m'}}\right)^{m'-1} \,
\langle\hat{P}_{f_1\ldots f_{m'}}\rangle\,
\hat{P}^{\mu\nu}_{qf_{m'+1}\ldots f_m}\,,
\eeq
where $\langle\hat{P}_{f_1\ldots f_{m'}}\rangle$ denotes the unpolarised splitting amplitude,
\beq
\langle\hat{P}_{f_1\ldots f_m}\rangle \equiv \frac{1}{N_{\textrm{pol}}}\,\delta_{hh'}\,\hat{P}^{hh'}_{f_1\ldots f_m}\,,
\eeq
and $N_{\textrm{pol}}$ denotes the number of physical polarisation states for the parent parton. We work in conventional dimensional regularisation (CDR), where the quarks and gluons have $2$ and $(D-2)$ polarisation states, respectively.

If also the sub-parent is a gluon, we can use a similar argument to that of Sect.~\ref{sec:collinear} to trade in the helicity indices $(h,h')$ for Lorentz indices $(\alpha,\beta)$ by amputating external polarisation vectors and contracting with polarisation tensors. 
Since only transverse polarisations are physical, only the transverse part of a Lorentz tensor carries physical information.
%This is equivalent to a projection onto the subspace transverse to the space spanned by $\pM{}$ and $n$, which we denote by $(\pM{},n)_\perp$. 
Thus, we can write eq.~\eqref{eq:P_coll_g} in the equivalent form,
\iffalse
%
\beq\label{eq:P_coll_g_g}
\mathscr{C}_{1\ldots m'} \hat{P}^{\mu\nu}_{f_1\ldots f_{m}}= \left(\frac{s_{[1\ldots m']\ldots m}}{s_{1\ldots m'}}\right)^{m'-1} \,
\hat{P}^{\alpha\beta}_{f_1\ldots f_{m'}}\,\hat{H}^{\alpha\beta;\mu\nu}_{f_{(1\ldots m')}f_{m'+1}\ldots f_m}\,,
\eeq
%
\fi
%
{\begin{align} \label{eq:P_coll_g_g}
\mathscr{C}_{(1\ldots m')\ldots m} & \mathscr{C}_{1\ldots m'} 
|\mathcal{M}_{f_1 \ldots f_n}(p_1,\ldots,p_n)|^2 = \left(\frac{2g_s^2\mu^{2\epsilon}}{s_{1\ldots m'}}\right)^{m'-1} \left(\frac{2g_s^2\mu^{2\epsilon}}{s_{[1\ldots m']\ldots m}}\right)^{m-m'} \nonumber\\
&\times 
\hat{P}_{f_1\ldots f_{m'},\alpha\beta}\,\hat{H}^{\alpha\beta;\mu\nu}_{f_{(1\ldots m')}f_{m'+1}\ldots f_m}\,\mathcal{T}_{ff_{m+1}\ldots f_n,\mu\nu}(\pM{},p_{m'+1},\ldots,p_n)\,,
\end{align}}
and the relation in eq.~\eqref{eq:H_to_P_helicities} becomes
\begin{align}
\hat{H}^{\alpha\beta;\mu\nu}_{f_{(1\ldots m')}f_{m'+1}\ldots f_m}
d_{\alpha \beta}(\pN{},n)
=\hat{P}^{\mu\nu}_{f_{(1\ldots m')}f_{m'+1}\ldots f_m}
+\,\text{gauge terms}\,.
\label{eq:HtoP_Lorentz}
\end{align}

In App.~\ref{app:tensor_struc} we show that the most general tensor structure of the splitting tensor is
\iffalse
%
\begin{align}
\hat{H}^{\alpha\beta;\mu\nu}_{gf_{m'+1}\ldots f_m} 
%&=\frac{1}{2}\left(\hat{H}^{\alpha\beta;\mu\nu}_{gf_{m'+1}\ldots f_m} +\hat{H}^{\beta\alpha;\mu\nu}_{gf_{m'+1}\ldots f_m}\right) \nonumber \\
&=d^{\alpha\beta}(\pM{},n)\,g^{\mu\nu}\,A^{(g)}_{gf_{m'+1}\ldots f_m} + \sum_{i,j=m'+1}^m\,\frac{\K{i}^{\mu}\K{j}^{\nu}}{s_{[1\ldots m']\ldots m}}\,d^{\alpha\beta}(\pM{},n)\,B^{(g)}_{ij,gf_{m'+1}\ldots f_m} \nonumber \\
&+\sum_{i,j=m'+1}^m\,\frac{\K{i}^{\alpha}\K{j}^{\beta}}{s_{[1\ldots m']\ldots m}}\,g^{\mu\nu}\,C^{(g)}_{ij,gf_{m'+1}\ldots f_m} +\sum_{i,j,k,l=m'+1}^m\,\frac{\K{i}^{\mu}\K{j}^{\nu}\K{k}^{\alpha}\K{l}^{\beta}}{s_{[1\ldots m']\ldots m}^2}\,D^{(g)}_{ijkl,gf_{m'+1}\ldots f_m} \nonumber \\
&+\sum_{i,j=m'+1}^m\,\frac{g^{\alpha\mu}\,\K{i}^{\beta}\K{j}^{\nu}+g^{\beta\nu}\,\K{i}^{\alpha}\K{j}^{\mu}+g^{\alpha\nu}\,\K{i}^{\beta}\K{j}^{\mu}+g^{\beta\mu}\,\K{i}^{\alpha}\K{j}^{\nu}}{s_{[1\ldots m']\ldots m}}\,E^{(g)}_{ij,gf_{m'+1}\ldots f_m}\nonumber \\
&+ \left(g^{\alpha\mu}\,g^{\beta\nu}
+g^{\alpha\nu}\,g^{\beta\mu}\right)\,F^{(g)}_{gf_{m'+1}\ldots f_m}\,. \label{eq:H_tensor_symm} 
\end{align}
%
\fi
%
\begin{align}
\hat{H}&{}^{\alpha\beta;\mu\nu}_{gf_{m'+1}\ldots f_m} 
%&=\frac{1}{2}\left(\hat{H}^{\alpha\beta;\mu\nu}_{gf_{m'+1}\ldots f_m} +\hat{H}^{\beta\alpha;\mu\nu}_{gf_{m'+1}\ldots f_m}\right) \nonumber \\
=d^{\alpha\beta}(\pM{},n)\,g^{\mu\nu}\,A^{(g)}_{gf_{m'+1}\ldots f_m} + \sum_{i,j=m'+1}^m\,\frac{\K{i}^{\mu}\K{j}^{\nu}}{s_{[1\ldots m']\ldots m}}\,d^{\alpha\beta}(\pM{},n)\,B^{(g)}_{ij,gf_{m'+1}\ldots f_m} \nonumber \\
&+\sum_{i,j=m'+1}^m\,\frac{\K{i}^{\alpha}\K{j}^{\beta}}{s_{[1\ldots m']\ldots m}}\,g^{\mu\nu}\,C^{(g)}_{ij,gf_{m'+1}\ldots f_m} +\sum_{i,j,k,l=m'+1}^m\,\frac{\K{i}^{\mu}\K{j}^{\nu}\K{k}^{\alpha}\K{l}^{\beta}}{s_{[1\ldots m']\ldots m}^2}\,D^{(g)}_{ijkl,gf_{m'+1}\ldots f_m} \nonumber \\
&+\sum_{i,j=m'+1}^m\,\frac{g^{\alpha\mu}\,\K{i}^{\beta}\K{j}^{\nu}+g^{\beta\nu}\,\K{i}^{\alpha}\K{j}^{\mu}+g^{\alpha\nu}\,\K{i}^{\beta}\K{j}^{\mu}+g^{\beta\mu}\,\K{i}^{\alpha}\K{j}^{\nu}}{s_{[1\ldots m']\ldots m}}\,E^{(g)}_{ij,gf_{m'+1}\ldots f_m} \nonumber\\ &+\left(g^{\alpha\nu}\,g^{\beta\mu}+g^{\alpha\mu}\,g^{\beta\nu}\right)\,F^{(g)}_{gf_{m'+1}\ldots f_m}\,. \label{eq:H_tensor_symm} 
\end{align}
At this point we have to make some comments about eq.~\eqref{eq:H_tensor_symm}.
First, let us discuss the symmetry properties of the splitting tensor $\hat{H}{}^{\alpha\beta;\mu\nu}_{gf_{m'+1}\ldots f_m}$. From its definition in eq.~\eqref{eq:H_Sp_Def} it follows that the splitting tensor must be symmetric under the exchange $(\mu,\alpha)\leftrightarrow(\nu,\beta)$. In eq.~\eqref{eq:H_def} it is contracted with $\hat{P}^{\alpha\beta}$ and $\cT^{\mu\nu}$, which are symmetric tensors at tree level. Hence, only the part of $\hat{H}{}^{\alpha\beta;\mu\nu}_{gf_{m'+1}\ldots f_m}$ that is individually symmetric under $\mu\leftrightarrow\nu$ and $\alpha\leftrightarrow\beta$ enters the factorisation in eq.~\eqref{eq:H_def}, and we only present here the part of the splitting tensor with this enlarged symmetry. Second, we see that eq.~\eqref{eq:H_tensor_symm} involves a mixture of metric tensors $g^{\rho\sigma}$ and polarisation tensors $d^{\rho\sigma}(\pM{},n)$. The tensor structure given in app.~\ref{app:tensor_struc} involves only polarisation tensors, which would make the splitting tensor explicitly transverse. However, similar to the case of the splitting amplitude discussed in Sec.~\ref{sec:collinear} (cf.~eq.~\eqref{eq:P_tensor}), some of the gauge-dependent terms drop out in the contraction in eq.~\eqref{eq:P_coll_g_g}. In particular, we can perform the replacements,
\begin{align}
    d^{\rho\sigma}(\pM{},n) \leftrightarrow -g^{\rho\sigma}\,, \quad (\rho,\sigma) \in \{(\mu,\nu),(\alpha,\mu),(\beta,\nu),(\alpha,\nu),(\beta,\mu)\}\,.
    \label{eq:transverse_rpl}
\end{align}
For $(\rho,\sigma)=(\mu,\nu)$, the equivalence between the polarisation tensor and the metric tensor follows from eq.~\eqref{eq:transverse_T}, while the other cases follow from relations like
{\begin{equation}\begin{split} 
\hat{P}_{f_1\ldots f_{m'},\alpha\beta}&\,d^{\alpha\mu}(\pM{},n)\,d^{\beta\nu}(\pM{},n)\,\mathcal{T}_{ff_{m+1}\ldots f_n,\mu\nu}=\hat{P}_{f_1\ldots f_{m'},\alpha\beta}\,g^{\alpha\mu}\,g^{\beta\nu}\,\mathcal{T}_{ff_{m+1}\ldots f_n,\mu\nu}\,,\\
\hat{P}_{f_1\ldots f_{m'},\alpha\beta}&\,d^{\alpha\mu}(\pM{},n)\,k_{\perp i}^{\beta}k_{\perp j}^{\nu}\,\mathcal{T}_{ff_{m+1}\ldots f_n,\mu\nu}=-\hat{P}_{f_1\ldots f_{m'},\alpha\beta}\,g^{\alpha\mu}\,k_{\perp i}^{\beta}k_{\perp j}^{\nu}\,\mathcal{T}_{ff_{m+1}\ldots f_n,\mu\nu}\,.
\end{split}\end{equation}}

%%%%COMMENT OUT%%%%%%%
\iffalse
Somewhat surprisingly, we retain the freedom to write the lower-order splitting amplitude in eq.~\eqref{eq:P_coll_g_g} in an explicitly transverse form. Contracting the right-hand-side of eq.~\eqref{eq:H_tensor_symm} with gauge terms $\pN{}$ and $n$, we obtain
%
\begin{align}
    \frac{\pN{}_\alpha n_\beta+\pN{}_\beta n_\alpha}{n\cdot\pN{}}\,
    \hat{H}^{\alpha\beta;\mu\nu}_{gf_{m'+1}\ldots f_m}
    &=
    \sum_{i,j=m'+1}^m\,\frac{\K{j}^\mu n^\nu+\K{j}^\nu n^\mu}{n\cdot\pN{}}\,\K{i}\cdot \pN{}\,E^{(g)}_{ij,gf_{m'+1}\ldots f_m} \nonumber \\
    &+ \frac{\pN{\mu} n^\nu+\pN{\nu} n^\mu}{n\cdot\pN{}} \,F^{(g)}_{gf_{m'+1}\ldots f_m}\,.
\end{align}
%
Indeed, the right-hand-side vanishes when contracting with $\mathcal{T}_{ff_{m+1}\ldots f_n}^{\mu\nu}$ since it is transverse to the orthogonal subspace $(\pM{},n)_\perp$. This is because $\pN{}$ lies in the direction $\pM{}$ up to terms that are power suppressed in the strongly-ordered collinear limit (cf. App.~\ref{app:nested}).
\fi
%%%%%%END COMMENT%%%%%%%%%
We have checked that our results for the quadruple splitting amplitudes have the correct behaviour in all strongly-ordered collinear limits, i.e., they satisfy eq.~\eqref{eq:P_coll} for $m'=2$ and $3$. The strongly-ordered limit of the quadruple splitting amplitudes involves the splitting tensors with two or three collinear particles in the final state. The relevant splitting tensors for two collinear partons can be found in ref.~\cite{Somogyi:2005xz},
\iffalse
%
\beq\bsp \label{eq:Hg_ff}
\hat{H}^{hh';\mu \nu}_{\bar q q}\,& =\frac{1}{2}\delta^{hh'} \hat P^{\mu\nu}_{\bar q q}\,,\\
\hat{H}^{\alpha \beta;\mu\nu}_{g g} 
\,&=
2C_A \left[
\frac{1-z}{z} g^{\alpha \mu} g^{\beta \nu} + \frac{z}{1-z} g^{\mu\nu} \frac{\K{}^\alpha\K{}^\beta}{\K{}^2} 
- z(1-z)\frac{\K{}^\mu\K{}^\nu}{\K{}^2} d^{\alpha \beta}(\pM{},n)
\right]\,,
\esp\eeq
%
\fi
%
\begin{align} \label{eq:Hg_ff}
\hat{H}^{hh';\mu \nu}_{\bar q q}\,& =\frac{1}{2}\delta^{hh'} \hat P^{\mu\nu}_{\bar q q}\,,\\
\nonumber\hat{H}^{\alpha \beta;\mu\nu}_{g g} 
\,&=
2C_A \left[
\frac{1-z}{2z} \left(g^{\alpha \mu} g^{\beta \nu}+g^{\alpha \nu} g^{\beta \mu} \right) + \frac{z}{1-z} g^{\mu\nu} \frac{\K{}^\alpha\K{}^\beta}{\K{}^2} 
- z(1-z)\frac{\K{}^\mu\K{}^\nu}{\K{}^2} d^{\alpha \beta}(\pM{},n)
\right]\,,
\end{align}
where we set $k_{\perp} = k_{1\perp}=-k_{2\perp}$. Note that, compared to ref.~\cite{Somogyi:2005xz}, we express the splitting tensor $\hat{H}^{\alpha \beta;\mu\nu}_{g g}$ in a form that is individually symmetric in $(\mu,\nu)$ and $(\alpha,\beta)$. In addition, the two-parton splitting tensor $\hat{H}^{\alpha \beta;\mu\nu}_{g g}$ is special in that certain tensor structures do not appear. The coefficients $A^{(g)}_{gg}$ and $D^{(g)}_{3333,gg}$ are subleading in the collinear limit, while $E
^{(g)}_{33,gg}$ vanishes in the explicitly symmetric form \eqref{eq:H_tensor_symm}. The relevant splitting tensors for three collinear partons are new and are given in ref.~\cite{DelDuca:2019ggv} and in App.~\ref{app:3partonhelicity_tensor}.
%
%\rayan{a repetition of the above discussion, but maybe we would like to say a few words to that effect below eq.~\eqref{eq:Hg_ff}.}
%Note that we could re-write $\hat{H}^{\mu\nu,\alpha \beta}_{g g}$ in an explicitly transverse form by making replacing the $D$-dimensional Minkowski metric by the transverse projector on the $(D-2)$-dimensional subspace according to eq.~\eqref{eq:transverse_rpl}.However, this is again completely redundant, since the extra gauge terms vanish upon contraction with the gauge independent helicity tensor $\mathcal T^{\mu\nu}_{f,f_{m'+1}\ldots f_m}$. On the other hand, the gauge terms in $d^{\alpha \beta}(\pM{},n)$ lead to non-trivial contributions to the strongly-ordered splitting amplitude when contracted with the lower order splitting amplitude $\hat P^{\alpha\beta}_{f_1f_2}$, $(f_1,f_2) \in \{(\bar q,q),(g,g)\}$.

\section{Conclusions}
\label{sec:conclusion}
In this paper, we have computed
the quadruple-collinear splitting amplitudes for a gluon parent in CDR. Combined with our previous results for a quark parent~\cite{DelDuca:2019ggv}, this completes the set of tree-level splitting amplitudes describing all collinear singularities for the emission of up to four collinear partons at N$^3$LO.
Our results are available in computer-readable form online~\cite{QuadColKernelsWebsite}. We have also considered the strongly-ordered limit when a subset of the four collinear partons become  collinear to each other, and we have derived the corresponding factorisation formul\ae. 
Our results satisfy the expected factorisations in all strongly-ordered limits, which provides a strong check on the correctness of our computations.

Our results are an important building block towards understanding the complete infrared structure of massless QCD amplitudes at N$^3$LO, which is a cornerstone to construct a substraction method at this order. Indeed,  firstly, the purely virtual infrared singularities of massless amplitudes with up to three loops are completely known~\cite{Catani:1998bh,Sterman:2002qn,Becher:2009cu,Gardi:2009qi,Almelid:2015jia,Almelid:2017qju}. 
Secondly, when our results are combined with the results for the one-loop emission of up to three collinear particles~\cite{Campbell:1997hg,Catani:1999ss,DelDuca:1999iql,Kosower:2002su,Bern:1994zx,Bern:1998sc,Kosower:1999rx,Bern:1999ry,Catani:2003vu,Badger:2015cxa}\footnote{But for the one-loop collinear splitting amplitude $q\to ggq$, which at present is unknown.} and the two-loop splitting amplitudes for two collinear partons~\cite{Bern:2004cz,Badger:2004uk,Duhr:2014nda}, they provide a complete description of all collinear singularities up to N$^3$LO. Finally, soft emissions are known for the tree-level emission of up to three soft partons~\cite{Bassetto:1984ik,Proceedings:1992fla,Catani:1999ss,Catani:2019nqv}, and at one and two loops for the emission of a single soft gluon~\cite{Bern:1998sc,Bern:1999ry,Catani:2000pi,Duhr:2013msa,Li:2013lsa, Dixon:2019lnw}. The soft current describing the emission of a pair of two soft partons at one-loop, however, is still missing. For the  future, it would be interesting to compute this current and to complete the description of all infrared singularities of massless QCD amplitudes at N$^3$LO.

    \section*{Acknowledgements}
CD and MM acknowledge the hospitality of the ETH Zurich, and MM also acknowledges the hospitality of the TH Department of CERN, at various stages of this work.
    This work was supported in part by the ERC starting grant  637019 ``MathAm'' (CD), the FRIA grant of the Fonds National de la Recherche Scientifique (FNRS), Belgium (MM), the European Research Council (ERC) under grant agreement No 694712 (PertQCD) and the Swiss National Science Foundation (SNF) under contract agreement No 179016 (VD, RH, AL).

	%%%%%%%%%%%%%%%%%%%%%%%%%%%%%%

	\appendix

\section{The iterated collinear limit}
\label{app:iterated_limit_of_split}

The strongly-ordered amplitude \cite{DelDuca:2019ggv},
\begin{equation}
\hat{P}_{f_1\ldots f_m}^{\,\text{s.o.}\,;ss'} = 
\hat{P}^{hh'}_{f_1\ldots f_{m'}}\,
\hat{H}^{hh';ss'}_{f_{(1\ldots m')}f_{m'+1}\ldots f_m}\,,
\label{eq:Pso}
\end{equation}
depends on the quantum numbers and light-cone kinematics of both the $m$-parton collinear set and its $m'$-parton subset. It is obtained by summing over the helicities $(h,h')$ of the parent parton of the collinear subset. In case the sub-parent with helicities $(h,h')$ is a quark, following the factorisation in eq.~\eqref{eq:Hg_qf}, the strongly-ordered splitting amplitude has the same tensor structure as an ordinary splitting amplitude. In case both the parent with helicities $(s,s')$ and the sub-parent are gluons, the strongly-ordered amplitude has a similar tensor structure as that of a gluon splitting amplitude,
\begin{align}
\hat{P}_{f_1\ldots f_m}^{\,\text{s.o.}\,;\mu\nu} =
g^{\mu\nu}\,A^{\,\text{s.o.}}_{f_1\ldots f_m} 
+ \sum_{i,j=1}^{m'}\, \frac{\Kb{i}^\mu \Kb{j}^\nu}{s_{1\ldots m'}} \, B_{ij,f_1\ldots f_m}^{\,\text{s.o.}}
+ \sum_{k,l=m'+1}^{m}\,\frac{\K{k}^\mu \K{l}^\nu}{s_{m'+1 \ldots m}} \, C_{kl,f_1\ldots f_m}^{\,\text{s.o.}}\,.
\end{align}
It depends on the transverse momenta $\Kb{i}^\mu$ of the collinear subset, $i=1,\ldots,m'$ and those of the other $(m-m')$-collinear partons unaffected by the strongly-ordered limit, $\K{j}^\mu$ with $j=m'+1,\ldots,m$. 

The strongly-ordered splitting amplitude can be obtained by performing the $m'$-parton iterated limit on the $m$-parton splitting amplitude,
\begin{equation}
 \hat{P}_{f_1\ldots f_m}^{\,\text{s.o.}\,;\mu\nu}
 %(\{\zeta_i,\Kb{i}\};\{z_j,k_j\};\epsilon) 
 =
 \left(\frac{s_{1\ldots m'}}{s_{[1\ldots m']\ldots m}}\right)^{m'-1}
\mathscr{C}_{1\ldots m'} \hat{P}^{\mu\nu}_{f_1\ldots f_m}
%(\{z_i,\K{i}\};\epsilon )
\,.
\label{eq:so_from_split}
\end{equation} 
The steps involved are essentially the same as for a quark parent, which was considered in ref.~\cite{DelDuca:2019ggv}. However, as a first step, we need to apply the relation
\begin{align}
\K{i}^\mu = \zeta_i K^\mu + \Kb{i}^\mu +  \frac{K\cdot \Kb{i}}{\alpha}\frac{ n^\mu}{ n \cdot \pM{}}\,, \quad i=1,\ldots,m' \,,
\end{align}
where $K^\mu = \sum_{j=1}^{m'} \,\K{j}^\mu = - \sum_{j=m' + 1}^m \, \K{j}^\mu$ and $\alpha = \sum_{j=1}^{m'}\, z_j$. Then, we perform the steps outlined in App.~C of ref.~\cite{DelDuca:2019ggv}.
%%%%%%%COMMENT OUT%%%%%%%%
%%%%%%%%%%%%%%%%%%%%%%%%%%

\iffalse
Next, we perform a light-cone decomposition of the sub-energies,
%
\begin{align}
\label{eq:so_sij_param2_1}
s_{ij} &\to \lambda'^2 s_{ij},\quad 1\leq i,j \leq m', \\
s_{ir} &\to
\zeta_i s_{[1\ldots m']r} +
\lambda'  \left( 2\K{r} \cdot \Kb{i} - \frac{2z_r}{\alpha}K \cdot \Kb{i}\right)
- \lambda'^2 \frac{ z_r}{\alpha \zeta_i}\Kb{i}^2, \label{eq:so_sij_param2_2}\\
&\quad 1\leq i\leq m', \quad m'+1 \leq r \leq m, \nonumber
\end{align}
%
where 
%
\begin{align}
s_{[1\ldots m']r} \equiv 2p_r \cdot \pN{} = - \frac{\alpha}{z_r} \, \K{r}^2 - 2\sum_{i=m' + 1}^m\, \K{i} \cdot \K{r} - \frac{z_r}{\alpha}\, K^2
\,.
\end{align}
%
The substitution \eqref{eq:so_sij_param2_2} was obtained by applying the non-trivial relation,
\begin{equation}
p_r \cdot \Kb{i} = \K{r} \cdot \Kb{i} - \frac{z_r}{\alpha} K \cdot \Kb{i},\quad 1\leq i\leq m', \quad m'+1 \leq r \leq m,
\end{equation}
where $K = \sum_{l=1}^{m'} \K{l}$.

Then, we need to apply the change of variables,
%
\begin{align}
 z_i = \zeta_i\left(1-\sum_{j=m'+1}^m z_j\right), \quad  1\leq i\leq m'\,,
\end{align}
%
in order to separate the kinematics of the lower-order $m'$-parton splitting process from that of the remaining $(m-m')$ partons with momenta $\{p_{m'+1},\ldots,p_m\}$. Finally, we apply the uniform rescaling $\Kb{i} \to \lambda^\prime \,\Kb{i}$ and determine the leading contribution to the Laurent series around $\lambda^\prime = 0$. 
\fi
%%%%%%%END COMMENT%%%%%%%%
%%%%%%%%%%%%%%%%%%%%%%%%%%

Comparing the strongly-ordered limit obtained from eq.~\eqref{eq:so_from_split} with that obtained from eq.~\eqref{eq:Pso}, works as a strong check on the splitting amplitudes. This was done to check all possible two- and three-parton sub-limits.

\section{Tensor structure of gluon parent and sub-parent splitting tensors}
\label{app:tensor_struc}

%\ClaudeComment{I agree with Rayan's comment below: the derivation must be made clearer, along the lines of Catani\&Grazzini: we should start from a squared off-shell current, with one final state parton index not summed over, and then repeat the discussion of Catani\&Grazzini. We should also make a distinction between two things: how far you get by looking at the splitting tensor extracted in this way, and constraints on the tensor structure because we assume it contracted with a splitting amplitude.}

In this appendix, we detail the general tensor structure of the gluon-initiated splitting tensor $\hat{H}^{\alpha \beta;\mu \nu}_{gf_{m'+1}\ldots f_m}$ with a gluon sub-parent, defined in Sec.~\ref{sec:nested_collinear}. This tensor appears as the collinear limit of the helicity tensor $\mathcal{T}^{\alpha\beta}$, defined in eq.~\eqref{eq:T_def}. Therefore, we begin by considering an $n$-parton tree-level QCD process with its associated helicity tensor $\mathcal{T}^{\alpha\beta}_{f_1 \ldots f_n}(p_1, \ldots, p_n)$, where parton $1$ carries the Lorentz indices $(\alpha,\beta)$.
%where the indices $\alpha$ and $\beta$ are the spin indices of the parton $1$, which is a gluon. 
Using the same kind of argument as for eq.~\eqref{eq:V_fac2}, our gauge choice allows us to consider only diagrams that have a propagator carrying the momentum $P= \sum_{i=1}^m p_i$ of the parent gluon.
%, which is a gluon in the case we are interested in. 
This directly implies that the sub-diagrams splitting the parent parton into $m$ partons will factorise (cf. eq.~\eqref{eq:V_fac2}), the only difference being that the spin indices of parton $1$ are not summed over,
\begin{align}\label{eq:H:fact}
    \mathscr{C}_{1 \ldots m}\,\mathcal{T}^{\alpha \beta}_{f_1 \ldots f_n}&(p_1, \ldots, p_n) = \\
    &\mathscr{C}_{1 \ldots m}\, \left[ \left( \frac{2 \mu^{2\eps}g_s^2}{s_{1\ldots m}} \right)^{m-1}
    \left[ \mathcal{M}^{(n)\,\mu}_{g f_{m+1} \ldots f_n} \right]^*  
    \,H^{(n)\,\alpha \beta;\mu\nu}_{f_1 \ldots f_m}(p_1,\ldots,p_m)\, 
    \mathcal{M}^{(n)\,\nu}_{g f_{m+1} \ldots f_n}\right]\,. \nonumber
\end{align}
The squared off-shell current $H^{(n)\,\alpha\beta;\mu\nu}_{f_1\ldots f_m}$ is the interference of the colour-dressed off-shell currents splitting the parent parton into $m$ collinear partons (cf. eq.~\eqref{eq:V_to_j}), with the Lorentz indices of parton $1$ and the parent parton open,
\begin{align}
\label{eq:T_to_j}
\left(\frac{2\mu^{2\epsilon}g_s^2}{s_{1\ldots m}}\right)^{m-1}\, H^{(n)\,\alpha \beta;\mu\nu }_{f_1\ldots f_m} =
\frac1{\mathcal{C}_{f}} \sum_{\substack{(s_2,\ldots,s_m)\\(c,c_1,\ldots,c_m)}}\,
\left[{\mathrm J}_{f_1\ldots f_m}^{c,c_1\ldots c_m;\nu \beta s_2 \ldots s_m}\right]^{\ast}\,
{\mathrm J}_{f_1\ldots f_m}^{c,c_1\ldots c_m;\mu \alpha s_2\ldots s_m}\,,
\end{align}
where we have suppressed the dependence on the momenta.
As usual the superscript $(n)$ indicates the dependence on the reference vector. Given the definition in eq.~\eqref{eq:T_to_j}, $H^{(n)\,\alpha \beta;\mu\nu }_{f_1\ldots f_m}$ is symmetric under the simultaneous exchange of the Lorentz indices of parton $1$ and the parent parton,
\begin{align}
\label{eq:Hn_symmetry}
H^{(n)\,\alpha\beta;\mu\nu}_{f_1 \ldots f_m}=H^{(n)\,\beta\alpha;\nu\mu}_{f_1 \ldots f_m}\,.
\end{align}
%
%Considering $\hat{H}^{(n)\,\alpha\beta;\mu\nu}_{f_1 \ldots f_m}$ as the sum of the interferences of the corresponding Feynman diagrams, 
%
%We write down the most general tensor structure for it, which can depend on all the collinear momenta, 
This means that the most general tensor structure for $H^{(n)\,\alpha \beta;\mu\nu }_{f_1\ldots f_m}$ is given by
\iffalse
%
\begin{multline}\label{eq:H_gen_first}
\hat{H}^{(n)\,\alpha \beta;\mu \nu}_{f_1 \ldots f_m}=A~g^{\mu\nu}g^{\alpha\beta}~+~B_{ij}~\frac{g^{\alpha\beta}p_i^{\mu}p_j^{\nu}}{s_{1\ldots m}}+~C_{ij}~\frac{g^{\mu\nu}p_i^{\alpha}p_j^{\beta}}{s_{1\ldots m}}~+~D_{ijkl}~\frac{p_i^{\mu}p_j^{\nu}p_k^{\alpha}p_l^{\beta}}{s_{1\ldots m}^2}\\
~~~~~~~~~~~+~E_{ij}~\frac{g^{\mu\alpha}p_i^{\nu}p_j^{\beta}+g^{\nu\beta}p_i^{\mu}p_j^{\alpha}+g^{\mu\beta}p_i^{\nu}p_j^{\alpha}+g^{\nu\alpha}p_i^{\mu}p_j^{\beta}}{s_{1\ldots m}}\\
~~~~~~~~~~~~~+~G_{ij}~\frac{g^{\mu\alpha}p_i^{\nu}p_j^{\beta}+g^{\nu\beta}p_i^{\mu}p_j^{\alpha}-g^{\mu\beta}p_i^{\nu}p_j^{\alpha}-g^{\nu\alpha}p_i^{\mu}p_j^{\beta}}{s_{1\ldots m}}\\
~+~F~\left(g^{\mu\alpha}g^{\nu\beta}+g^{\mu\beta}g^{\nu\alpha}\right)
~+~J~\left(g^{\mu\alpha}g^{\nu\beta}-g^{\mu\beta}g^{\nu\alpha}\right)
~+~\text{$n$-terms},
\end{multline}
%
%%%%%%%%
\rayan{I think the above is wrong. It should only have the block symmetry. It should be:}\rayan{sorry Martin, I see what you mean.}
\fi
%
\begin{align}
H^{(n)\,\alpha\beta;\mu\nu}_{gf_1\ldots f_m} 
&=g^{\alpha\beta}\,g^{\mu\nu}\,\overline{A}^{(g)}_{g f_1\ldots f_m} + \sum_{i,j=1}^m\,\frac{p_i^{\mu} p_j^{\nu}}{s_{1\ldots m}}\,g^{\alpha\beta}\,\overline{B}^{(g)}_{ij,gf_1\ldots f_m} 
+\sum_{i,j=1}^m\,\frac{p_i^{\alpha}p_j^{\beta}}{s_{1\ldots m}}\,g^{\mu\nu}\,\overline{C}^{(g)}_{ij,gf_1\ldots f_m} \nonumber\\
&+\sum_{i,j,k,l=1}^m\,\frac{p_i^{\mu}p_j^{\nu}p_k^{\alpha}p_l^{\beta}}{s_{1\ldots m}^2}\,\overline{D}^{(g)}_{ijkl,gf_{1}\ldots f_m} 
+\sum_{i,j=1}^m\,\frac{g^{\alpha\mu}\,p_i^{\beta}p_j^{\nu}+g^{\beta\nu}\,p_i^{\alpha}p_j^{\mu}}{s_{1\ldots m}}\,\overline{E}^{(g)}_{1\,ij,gf_{1}\ldots f_m} \nonumber \\
&+\sum_{i,j=1}^m\,\frac{g^{\alpha \nu}\,p_i^{\beta}p_j^{\mu}+g^{\beta\mu}\,p_i^{\alpha}p_j^{\nu}}{s_{1\ldots m}}\,\overline{E}^{(g)}_{2\,ij,gf_{m'+1}\ldots f_m}
+ g^{\alpha\mu}\,g^{\beta\nu}\,\overline{F}^{(g)}_{1\,gf_{1}\ldots f_m} \nonumber \\
&+g^{\alpha\nu}\,g^{\beta\mu}\,\overline{F}^{(g)}_{2\,gf_{1}\ldots f_m}
+ \text{ gauge terms}
\,,
\label{eq:H_n_tensor} 
\end{align}
%
%where all repeated parton indices in the right-hand side are summed over the set of collinear partons, and `$n$-terms' denotes terms that are proportional to $n$. By the same arguments as for the structure of the splitting amplitude for the gluon parent, the coefficients $A$, $B_{ij}$, $C_{ij}$, $D_{ijkl}$, $E_{ij}$, $F$, $G_{ij}$, and $J$ are dimensionless and of order zero in the collinear limit. In addition, we have the following symmetries, which follow directly from eq.~\eqref{eq:Hn_symmetry},
where {`gauge terms'} on the right-hand-side denote all terms proportional to the axial gauge vector $n$. By the same arguments as for the tensor structure of the gluon-initiated splitting amplitude, the coefficients $\overline A^{(g)}_{gf_1\ldots f_m}$ through $\overline F^{(g)}_{2\,gf_1\ldots f_m}$ in eq.~\eqref{eq:H_n_tensor} are dimensionless and of order zero in the collinear limit. In addition, we have the following symmetries, which follow directly from eq.~\eqref{eq:Hn_symmetry},
\begin{equation}\bsp
    \overline{B}^{(g)}_{ij,gf_{1}\ldots f_m} &= \overline{B}^{(g)}_{ji,gf_{1}\ldots f_m}\,,  \\
    \overline{C}^{(g)}_{ij,gf_{1}\ldots f_m} &= \overline{C}^{(g)}_{ji,gf_{1}\ldots f_m}\,,  \\
    \overline{D}^{(g)}_{ijkl,gf_{1}\ldots f_m} &= \overline{D}^{(g)}_{jilk,gf_{1}\ldots f_m}\,.
\esp\end{equation}
%
%Next, to make sure that this tensor is transverse, we project it by use of spin-polarisation tensors on both the parent parton and parton $1$ legs,
Since only the transverse part of a Lorentz tensor holds physical information (cf. Secs.~\ref{sec:collinear} and \ref{sec:nested_collinear}), we multiply $H^{(n)\,\alpha\beta;\mu\nu}_{gf_1\ldots f_m}$ by spin-polarisation tensors,
\begin{align}
d_{\mu \tilde{\mu}}(P,n)\,d_{\nu \tilde{\nu}}(P,n)\,d_{\alpha \tilde{\alpha}}(p_1,n)\,d_{\beta \tilde{\beta}}(p_1,n)\,H^{(n)\,\tilde{\alpha} \tilde{\beta};\tilde{\mu} \tilde{\nu}}_{f_1\ldots f_m}\,.
\label{eq:H_n_project}
\end{align}
%
%Inserting the right-hand-side of eq.~\eqref{eq:H_n_tensor} into eq.~\eqref{H_n_project}, the explicit results of the contractions blabla
Next, we take the $m$-parton collinear limit of $H^{(n)\,\alpha\beta;\mu\nu}_{ f_1\ldots f_m}$ to obtain the splitting tensor. Equation~\eqref{eq:H:fact} can then be written as
\begin{equation}\label{eq:H:coll_limit}
    \mathscr{C}_{1 \ldots m}\,
    \mathcal{T}^{\alpha \beta}_{f_1 \ldots f_n}(p_1, \ldots, p_n)  
    = \left( \frac{2 \mu^{2\eps}g_s^2}{s_{1\ldots m}} \right)^{m-1}
      \hat{H}^{\alpha\beta;\mu\nu}_{f_1 \ldots f_m}\, \mathcal{T}_{g f_{m+1} \ldots f_n,\mu \nu}(\pM{},p_{m+1}, \ldots, p_n).
\end{equation}
By computing explicitly the contractions of the tensors on the right-hand-side of eq.~\eqref{eq:H_n_tensor} with the spin-polarisation tensors, we obtain
\begin{align}
d_{\mu\alpha}(p_i,n)\,g^{\alpha\beta}\,d_{\nu\beta}(p_i,n)&=
-d_{\mu\nu}(\pM{},n)+\ldots\,, \nonumber \\
{d^{\mu}}_\nu(p_j,n)\,p_i^{\nu}&=
-\K{i}^{\mu}+\frac{z_i}{z_j}\K{j}^{\mu}+\ldots\,, \nonumber \\
d^{\mu\nu}(P,n)\,P_{\nu}&=0 + \ldots\,,\nonumber \\ d^{\mu\nu}(P,n)\,n_{\nu}&=0\,, \label{eq:H_n_contract} \\
{d^{\mu}}_\nu(p_i,n)\,\pM{\nu}&=
\frac{1}{z_i}\,\K{i}^{\mu}+\ldots\,, \nonumber \\
{d^{\mu}}_\nu(\pM{},n)\,p_i^\nu&=
-\K{i}^{\mu}+\ldots\,, \nonumber
\end{align}
where in the ellipses we have suppressed sub-leading terms in the $m$-parton collinear limit. 
%The terms proportional to the reference vector $n$ in eq.~\eqref{eq:H_n_tensor} vanish since they are transverse to the spin-polarisation tensors. 
Thus, the transverse part of the splitting tensor in eq.~\eqref{eq:H:coll_limit} contains only combinations of spin-polarisation tensors and transverse momenta. 

%%%%%%%COMMENT OUT%%%%%%%
%%%%%%%%%%%%%%%%%%%%%%%%%
\iffalse
%
\begin{align}
d_{\mu\alpha}(p_1,n)g^{\alpha\beta}d_{\nu\beta}(p_1,n)&=-d_{\mu\nu}(\tilde{P},n)+\ldots, \nonumber \\
d^{\mu\nu}(p_1,n)p_{i,\nu}&=-\tilde{k}_{\perp i}^{\mu}-\frac{z_i}{z_1} \tilde{k}_{\perp 1}^{\mu}+\ldots, \nonumber \\
d^{\mu\nu}(P,n)P_{\nu}&=0+\ldots, \nonumber \\
d^{\mu\nu}(P,n)p_{1\nu}&=\tilde{k}_{\perp 1}^{\mu}+\ldots, \nonumber \\
d^{\mu\nu}(p_1,n)P_{\nu}&=\frac{1}{z_1}~\tilde{k}_{\perp 1}^{\mu}+\ldots, \nonumber \\
d^{\mu\nu}(P,n)p_{i,\nu}&=-\tilde{k}_{\perp i}^{\mu}+\ldots,
\end{align}
where the dots stand for higher order terms in the collinear limit, one realises that the only possible tensor structures for the transverse splitting tensor are combinations of spin-polarisation tensors and $\tilde{k}_{\perp i}$. In addition, we can take into account the fact that the splitting tensor is contracted with $\mathcal{T}^{\mu \nu}_{g f_{m+1} \ldots f_n}$ in eq.~\eqref{eq:H:coll_limit}, which is symmetric in $\mu \leftrightarrow \nu$, which allows us to discard all anti-symmetric terms in $\mu \leftrightarrow \nu$. This leads to the following expression for the splitting tensor,
\fi
%%%%%%%END COMMENT%%%%%%%
%%%%%%%%%%%%%%%%%%%%%%%%%

In addition, we can take into account that in eq.~\eqref{eq:H:coll_limit} the splitting tensor is contracted with the symmetric tensor $\mathcal{T}^{\mu \nu}_{g f_{m+1} \ldots f_n}$, which allows us to discard terms in $\hat{H}^{\alpha\beta;\mu\nu}_{f_1 \ldots f_m}$ that are anti-symmetric under the exchange $\mu \leftrightarrow \nu$. This leads to the following expression for the splitting tensor,
\begin{align}
\hat H^{\,\alpha\beta;\mu\nu}_{gf_1\ldots f_m} 
&=-d^{\alpha\beta}\,d^{\mu\nu}\,A^{(g)}_{g f_1\ldots f_m} + \sum_{i,j=1}^m\,\frac{\K{i}^{\mu} \K{j}^{\nu}}{s_{1\ldots m}}\,d^{\alpha\beta}\,B^{(g)}_{ij,gf_1\ldots f_m} \nonumber \\
&-\sum_{i,j=1}^m\,\frac{\K{i}^{\alpha}\K{j}^{\beta}}{s_{1\ldots m}}\,d^{\mu\nu}\,C^{(g)}_{ij,gf_1\ldots f_m}
+\sum_{i,j,k,l=1}^m\,\frac{\K{i}^{\mu}\K{j}^{\nu}\K{k}^{\alpha}\K{l}^{\beta}}{s_{1\ldots m}^2}\, D^{(g)}_{ijkl,gf_{1}\ldots f_m} \nonumber \\
&-\sum_{i,j=1}^m\,\frac{
d^{\alpha\mu}\,\K{i}^{\beta}\K{j}^{\nu}
+d^{\beta\nu}\,\K{i}^{\alpha}\K{j}^{\mu}
+d^{\alpha \nu}\,\K{i}^{\beta}\K{j}^{\mu}
+d^{\beta\mu}\,\K{i}^{\alpha}\K{j}^{\nu}}{s_{1\ldots m}}\,E^{(g)}_{\,ij,gf_{1}\ldots f_m} \nonumber \\
&+ \left(d^{\alpha\mu}\,d^{\beta\nu}+d^{\alpha\nu}\,d^{\beta\mu}\right)\,F^{(g)}_{gf_{1}\ldots f_m} 
\,,
\label{eq:H_sym_munu} 
\end{align}
where we have suppressed the dependence of the spin-polarisation tensor on the gauge vectors $\pM{}$ and $n$. The scalar coefficients appearing in eq.~\eqref{eq:H_sym_munu} are linear combinations of the coefficients in eq.~\eqref{eq:H_n_tensor}.
To keep the symmetry of this expression, the coefficients now satisfy
\iffalse
%
\begin{align}
    \bar{B}_{ij} &= \bar{B}_{ji}, \nonumber \\
    \bar{C}_{ij} &= \bar{C}_{ji}, \nonumber \\
    \bar{D}_{ijkl} = \bar{D}_{jikl} &= \bar{D}_{ijlk}=\bar{D}_{jilk}.
\end{align}
\fi
%
%
\begin{equation}\bsp
    B^{(g)}_{ij,gf_1\ldots f_m} &= B^{(g)}_{ji,gf_1\ldots f_m}\,, \\
    C^{(g)}_{ij,gf_1\ldots f_m} &= C^{(g)}_{ji,gf_1\ldots f_m}\,,  \\
    D^{(g)}_{ijkl,gf_1\ldots f_m} = D^{(g)}_{jikl,gf_1\ldots f_m} &= D^{(g)}_{ijlk,gf_1\ldots f_m}=D^{(g)}_{jilk,gf_1\ldots f_m}\,. 
\esp\end{equation}
We can now immediately obtain eq.~\eqref{eq:H_tensor_symm} from eq.~\eqref{eq:H_sym_munu} by replacing some of the polarisation tensors by metric tensors following the discussion in Sec.~\ref{sec:nested_collinear}.

\section{The three-parton splitting tensors}
\label{app:3partonhelicity_tensor}
There are three gluon-initiated splitting tensors with three partons in the collinear set,
\begin{align}
    g \to \bar q q g, \quad g \to g \bar q q, \quad g\to g g g\,.
    \label{eq:split_tensors}
\end{align}
The first is obtained by not summing over the helicities $(h,h')$ of the quark (or equivalently, by charge-conjugation, the anti-quark) in the collinear set. It is proportional to the three-parton splitting amplitude $\hat{P}^{\mu\nu}_{\bar q q g}$ given in ref.~\cite{Catani:1999ss}, 
\begin{align}
\label{eq:Hg_qqg}
\hat{H}^{hh';\mu\nu}_{\bar q q g}&= \frac{1}{2}\,\delta^{hh'}\,\hat{P}^{\mu\nu}_{\bar q q g}\,.
\end{align}
The other two have the tensor structure of eq.~\eqref{eq:H_tensor_symm}. The three-gluon splitting tensor $\hat{H}^{\alpha\beta,\mu\nu}_{g g g}$ is too lengthy to be presented on paper, but we make it available in computer-readable form~\cite{QuadColKernelsWebsite}. In the remainder of this section, we provide explicit results for the splitting tensor $\hat{H}^{\alpha \beta;\mu\nu}_{g \bar q q}$. We can write it in terms of an `abelian' and `non-abelian' part,
\begin{align}
\label{eq:Hg_gqq}
\hat{H}^{\alpha \beta;\mu\nu}_{g \bar q q}&=
\frac{1}{2}\,C_F\, \hat{H}^{\alpha \beta,\mu\nu\text{ (ab)}}_{g \bar q q}
+\frac{1}{2}\,C_A\, \hat{H}^{\alpha \beta,\mu\nu\text{ (nab)}}_{g \bar q q}\,.
\end{align}

We add a subscript $(12)$ to denote the on-shell momentum of the gluon sub-parent. Furthermore, we define the shorthand,
\begin{equation}
    z_{1\ldots j} = z_1 + \ldots + z_j\,, \qquad
    \bar{z}_i = 1- z_i\,, \qquad
    \K{1\ldots j} = \K{1} + \ldots + \K{j}\,.
\end{equation}
In what follows, we have eliminated $z_{12}$ and $\K{12}$ using the constraints. The sub-energies $s_{[ij]k}$ are defined in eq.~\eqref{eq:s_square}. 
The coefficients in eq.~\eqref{eq:H_tensor_symm} belonging to the `abelian' piece of $\hat H^{\alpha\beta;\mu\nu}_{g_{(12)}\bar q_3 q_4}$ are given by
%%%%%%%%%%
%%%%%%%%%%
\begingroup
\allowdisplaybreaks
% TR CF piece
\begin{align}
A^{(g)\text{ (ab)}}_{g_{(12)} \bar q_3 q_4} &= -\frac{\left(s_{[12]3}+s_{[12]4}\right)^2}{2 s_{[12]3}s_{[12]4}}\,, \\
%%%%%%%%%%
%%%%%%%%%%
B_{33,g_{(12)} \bar q_3 q_4}^{(g)\text{ (ab)}} &\equiv 
B_{34,g_{(12)}\bar q_3 q_4}^{(g)\text{ (ab)}}
\equiv
B_{43,g_{(12)}\bar q_3 q_4}^{(g)\text{ (ab)}}
\equiv 
B_{44,g_{(12)}\bar q_3 q_4}^{(g)\text{ (ab)}}=-\frac{2s_{[12]34}}{s_{[12]3}s_{[12]4}}\,, \\
%%%%%%%%%
%%%%%%%%%
C_{33,g_{(12)} \bar q_3 q_4}^{(g)\text{ (ab)}} &=
-\frac{2s_{[12]34}\left(z_4s_{[12]3}-\bar{z}_4s_{[12]4}\right){}^2}{s_{[12]3}^2s_{[12]4}^2(1-z_{34})^2}
\,, \\
%%%%%%%%%%
%%%%%%%%%%
C_{44,g_{(12)}\bar q_3 q_4}^{(g)\text{ (ab)}}&=
-\frac{2s_{[12]34}\left(z_3s_{[12]4}-\bar{z}_3s_{[12]3}\right){}^2}{s_{[12]3}^2s_{[12]4}^2(1-z_{34})^2}
\,, \\
%%%%%%%%%%
%%%%%%%%%%
C_{34,g_{(12)}\bar q_3 q_4}^{(g)\text{ (ab)}}&\equiv C_{43,g_{(12)}\bar q_3 q_4}^{(g)\text{ (ab)}}=
\frac{2s_{[12]34}\left(z_3s_{[12]4}-\bar{z}_3s_{[12]3}\right)\left(z_4s_{[12]3}-\bar{z}_4s_{[12]4}\right)}{s_{[12]3}^2s_{[12]4}^2(1-z_{34})^2}
\,, \\
%%%%%%%%%
%%%%%%%%%
D_{3333,g_{(12)}\bar q_3 q_4}^{(g)\text{ (ab)}}&=
-\frac{8z_4{}^2}{s_{[12]4}^2(1-z_{34})^2}
\,, \\
%%%%%%%%%
%%%%%%%%%
D_{4444,g_{(12)}\bar q_3 q_4}^{(g)\text{ (ab)}}&=
-\frac{8z_3{}^2}{s_{[12]3}^2(1-z_{34})^2}
\,, \\
%%%%%%%%%
%%%%%%%%%
D_{3334,g_{(12)}\bar q_3 q_4}^{(g)\text{ (ab)}}&\equiv
D_{3343,g_{(12)}\bar q_3 q_4}^{(g)\text{ (ab)}}=
-\frac{8z_4\bar{z}_3}{s_{[12]4}^2(1-z_{34})^2}
\,, \\
%%%%%%%%%
%%%%%%%%%
D_{3433,g_{(12)}\bar q_3 q_4}^{(g)\text{ (ab)}}&\equiv
D_{4333,g_{(12)}\bar q_3 q_4}^{(g)\text{ (ab)}}=
-\frac{8z_4\bar{z}_4}{s_{[12]3}s_{[12]4}(1-z_{34})^2}
\,, \\
%%%%%%%%%
%%%%%%%%%
D_{4434,g_{(12)}\bar q_3 q_4}^{(g)\text{ (ab)}}&\equiv
D_{4443,g_{(12)}\bar q_3 q_4}^{(g)\text{ (ab)}}=
-\frac{8z_3\bar{z}_4}{s_{[12]3}^2(1-z_{34})^2}
\,, \\
%%%%%%%%%
%%%%%%%%%
D_{3444,g_{(12)}\bar q_3 q_4}^{(g)\text{ (ab)}}&\equiv
D_{4344,g_{(12)}\bar q_3 q_4}^{(g)\text{ (ab)}}=
-\frac{8z_3\bar{z}_3}{s_{[12]3}s_{[12]4}(1-z_{34})^2}
\,, \\
%%%%%%%%%
%%%%%%%%%
D_{3434,g_{(12)}\bar q_3 q_4}^{(g)\text{ (ab)}}&\equiv
D_{4343,g_{(12)}\bar q_3 q_4}^{(g)\text{ (ab)}} \equiv D_{4334,g_{(12)}\bar q_3 q_4}^{(g)\text{ (ab)}} \equiv 
D_{3443,g_{(12)}\bar q_3 q_4}^{(g)\text{ (ab)}} \nonumber \\
&=4\,\frac{z_3\bar{z}_4+z_4\bar{z}_3-1}{s_{[12]3}s_{[12]4}(1-z_{34})^2}
\,, \\
%%%%%%%%%
%%%%%%%%%
%%%%%%%%%
%%%%%%%%%
%D_{3434,g_{(12)}\bar q_3 q_4}^{(g)\text{ (ab)}}&\equiv
%D_{4343,g_{(12)}\bar q_3 q_4}^{(g)\text{ (ab)}} 
%=-\frac{2}{(1-z_{34})}\left(\frac{1}{s_{[12]3}^2}+\frac{1}{s_{[12]4}^2} \right) 
%\nonumber \\&\text{} \hspace{2.8cm}
%- 4\frac{\bar z_4 \bar z_3 + z_4 z_3}{(1-z_{34})^2 s_{[12]3}s_{[12]4}}
%\,, \\
%%%%%%%%%
%%%%%%%%%
%D_{4334,g_{(12)}\bar q_3 q_4}^{(g)\text{ (ab)}} &\equiv 
%D_{3443,g_{(12)}\bar q_3 q_4}^{(g)\text{ (ab)}}
%=\frac{2}{(1-z_{34})}\left(\frac{1}{s_{[12]3}^2}+\frac{1}{s_{[12]4}^2} \right) 
%\nonumber \\&\text{} \hspace{2.8cm}
%- 4\frac{\bar z_4 \bar z_3 + z_4 z_3}{(1-z_{34})^2 s_{[12]3}s_{[12]4}}
%\,, \\
%%%%%%%%%
%%%%%%%%%
D_{4433,g_{(12)}\bar q_3 q_4}^{(g)\text{ (ab)}}&=
-\frac{8\bar{z}_4^2}{s_{[12]3}^2(1-z_{34})^2}
\,, \\
%%%%%%%%%
%%%%%%%%%
D_{3344,g_{(12)}\bar q_3 q_4}^{(g)\text{ (ab)}}&=
-\frac{8\bar{z}_3^2}{s_{[12]4}^2(1-z_{34})^2}
\,, \\
%%%%%%%%%
%%%%%%%%%
E_{33,g_{(12)} \bar q_3 q_4}^{(g)\text{ (ab)}} &=
-\frac{2z_4}{s_{[12]4}(1-z_{34})}
\,, \\
%%%%%%%%%%
%%%%%%%%%%
E_{44,g_{(12)}\bar q_3 q_4}^{(g)\text{ (ab)}}&=
-\frac{2z_3}{s_{[12]3}(1-z_{34})}
\,, \\
%%%%%%%%%%
%%%%%%%%%%
E_{34,g_{(12)}\bar q_3 q_4}^{(g)\text{ (ab)}}&=
-\frac{2\bar{z}_4}{s_{[12]3}(1-z_{34})}
\,, \\
%%%%%%%%%%
%%%%%%%%%%
E_{43,g_{(12)}\bar q_3 q_4}^{(g)\text{ (ab)}}&=
-\frac{2\bar{z}_3}{s_{[12]4}(1-z_{34})}
\,, \\
%%%%%%%%%%
%%%%%%%%%%
F_{g_{(12)}\bar q_3 q_4}^{(g)\text{ (ab)}}&= -1\,,
\end{align}
%%%%%%%%%%%%%%%%%%%%%%%%%%%%%%%%%%%%%%%%%%%%%%%%%%%%%%%%%%
%%%%%%%%%%%%%%%%%%%%%%%%%%%%%%%%%%%%%%%%%%%%%%%%%%%%%%%%%%
%%%%%%%%%%%%%%%%%%%%%%%%%%%%%%%%%%%%%%%%%%%%%%%%%%%%%%%%%%
%%%%%%%%%%%%%%%%%%%%%%%%%%%%%%%%%%%%%%%%%%%%%%%%%%%%%%%%%%
%
while the `non-abelian' coefficients read
% TR*CA piece
\begin{align}
A^{(g)\text{ (nab)}}_{g_{(12)} \bar q_3 q_4} &= \frac{1}{2}
\,, \\
%%%%%%%%%%
%%%%%%%%%%
B_{33,g_{(12)} \bar q_3 q_4}^{(g)\text{ (nab)}} &=
\frac{s_{[12]34}\left(s_{34}+s_{[12]4}\right)}{s_{34}s_{[12]3}s_{[12]4}}\,,\\
%%%%%%%%%
%%%%%%%%%
B_{34,g_{(12)}\bar q_3 q_4}^{(g)\text{ (nab)}}&\equiv
B_{43,g_{(12)}\bar q_3 q_4}^{(g)\text{ (nab)}}=
\frac{s_{[12]34}\left(s_{[12]34}+s_{34}\right)}{2s_{34}s_{[12]3}s_{[12]4}}
\,, \\
%%%%%%%%%%
%%%%%%%%%%
B_{44,g_{(12)} \bar q_3 q_4}^{(g)\text{ (nab)}} &=
\frac{s_{[12]34}\left(s_{34}+s_{[12]3}\right)}{s_{34}s_{[12]3}s_{[12]4}}
\,, \\
%%%%%%%%%
%%%%%%%%%
C_{33,g_{(12)} \bar q_3 q_4}^{(g)\text{ (nab)}} &=
-\frac{s_{[12]34}\left(\bar{z}_4\left(2z_4s_{34}+s_{[12]4}\right)+z_4s_{[12]3}\right)}{s_{34}s_{[12]3}s_{[12]4}(1-z_{34})^2}
\,, \\
%%%%%%%%%%
%%%%%%%%%%
C_{44,g_{(12)}\bar q_3 q_4}^{(g)\text{ (nab)}}&=
-\frac{s_{[12]34}\left(\bar{z}_3\left(2z_3s_{34}+s_{[12]3}\right)+z_3s_{[12]4}\right)}{s_{34}s_{[12]3}s_{[12]4}(1-z_{34})^2}
\,, \\
%%%%%%%%%%
%%%%%%%%%%
C_{34,g_{(12)}\bar q_3 q_4}^{(g)\text{ (nab)}}&\equiv C_{43,g_{(12)}\bar q_3 q_4}^{(g)\text{ (nab)}} =-\frac{s_{[12]34}}{(1-z_{34})^2}
\Bigg(
\frac{\bar{z}_4\bar{z}_3 + z_4z_3}{s_{[12]3}s_{[12]4}}
+\frac{\bar{z}_3+z_4}{2s_{34}s_{[12]4}} +\frac{\bar{z}_4+z_3}{2s_{34}s_{[12]3}}
\Bigg)\nonumber
\,, \\
%%%%%%%%%
%%%%%%%%%
D_{3333,g_{(12)}\bar q_3 q_4}^{(g)\text{ (nab)}}&=
-\frac{8z_ 4{}^2\left(s_{34}+s_{[12]4}\right)}{s_{34}{}^2s_{[12]4}(1-z_{34})^2}
\,, \\
%%%%%%%%%
%%%%%%%%%
D_{4444,g_{(12)}\bar q_3 q_4}^{(g)\text{ (nab)}}&=
-\frac{8z_ 3{}^2\left(s_{34}+s_{[12]3}\right)}{s_{34}{}^2s_{[12]3}(1-z_{34})^2}
\,, \\
%%%%%%%%%
%%%%%%%%%
D_{3334,g_{(12)}\bar q_3 q_4}^{(g)\text{ (nab)}}&\equiv
D_{3343,g_{(12)}\bar q_3 q_4}^{(g)\text{ (nab)}}=
-\frac{8z_ 4\bar{z}_3\left(s_{34}+s_{[12]4}\right)}{s_{34}{}^2s_{[12]4}(1-z_{34})^2}
\,, \\
%%%%%%%%%
%%%%%%%%%
D_{3433,g_{(12)}\bar q_3 q_4}^{(g)\text{ (nab)}}&\equiv
D_{4333,g_{(12)}\bar q_3 q_4}^{(g)\text{ (nab)}}=
\frac{4z_ 4\bar{z}_4}{s_{[12]4}(1-z_{34})^2} \Bigg(
\frac{s_{34}+2s_{[12]4}}{s_{34}^2}
+\frac{s_{34}+s_{[12]4}}{s_{34}s_{[12]3}}
\Bigg)
\,, \\
%%%%%%%%%
%%%%%%%%%
D_{4434,g_{(12)}\bar q_3 q_4}^{(g)\text{ (nab)}}&\equiv
D_{4443,g_{(12)}\bar q_3 q_4}^{(g)\text{ (nab)}}=
-\frac{8z_ 3\bar{z}_ 4\left(s_{34}+s_{[12]3}\right)}{s_{34}{}^2s_{[12]3}(1-z_{34})^2}
\,, \\
%%%%%%%%%
%%%%%%%%%
D_{3444,g_{(12)}\bar q_3 q_4}^{(g)\text{ (nab)}}&\equiv
D_{4344,g_{(12)}\bar q_3 q_4}^{(g)\text{ (nab)}}=
\frac{4z_ 3\bar{z}_3}{s_{[12]4}(1-z_{34})^2} \Bigg(
\frac{s_{34}+2s_{[12]4}}{s_{34}^2}
+\frac{s_{34}+s_{[12]4}}{s_{34}s_{[12]3}}
\Bigg)
\,, \\
%%%%%%%%%
%%%%%%%%%
D_{3434,g_{(12)}\bar q_3 q_4}^{(g)\text{ (nab)}}&\equiv
D_{4343,g_{(12)}\bar q_3 q_4}^{(g)\text{ (nab)}} \equiv D_{4334,g_{(12)}\bar q_3 q_4}^{(g)\text{ (nab)}} \equiv 
D_{3443,g_{(12)}\bar q_3 q_4}^{(g)\text{ (nab)}}
\nonumber \\
&=2\,\frac{\bar{z}_4\bar{z}_3+z_3z_4}{(1-z_{34})^2}\,
\left(
\frac{2}{s_{34}^2}+ \frac{s_{[12]34}}{s_{[12]3}s_{[12]4}s_{34}}
\right)
\,, \\
%%%%%%%%%
%%%%%%%%%
%D_{3434,g_{(12)}\bar q_3 q_4}^{(g)\text{ (nab)}}&\equiv
%D_{4343,g_{(12)}\bar q_3 q_4}^{(g)\text{ (nab)}}
%=
%\frac{4z_3z_4}{s_{[12]4}(1-z_{34})^2} \Bigg(
%\frac{s_{34}+2s_{[12]4}}{s_{34}^2}
%+\frac{s_{34}+s_{[12]4}}{s_{34}s_{[12]3}}
%\Bigg) \\
%&\text{}\hspace{2.8cm}
%-\frac{2}{s_{[12]3}s_{[12]4}(1-z_{34})} \nonumber
%\,, \\
%%%%%%%%%
%%%%%%%%%
%D_{4334,g_{(12)}\bar q_3 q_4}^{(g)\text{ (nab)}}&\equiv
%D_{3443,g_{(12)}\bar q_3 q_4}^{(g)\text{ (nab)}} =\frac{2}{(1-z_{34})^2s_{34}s_{[12]4}} \Bigg(
%\frac{z_3z_4+\bar{z}_4\bar{z}_3(s_{34}+2s_{[12]4})}{s_{[12]3}} \\
%&\text{}\hspace{2.8cm}+\frac{2\bar{z}_4\bar{z}_3(s_{34}+2s_{[12]4})}{s_{34}}
%\Bigg)\nonumber
%\,, \\
%%%%%%%%%
%%%%%%%%%
D_{4433,g_{(12)}\bar q_3 q_4}^{(g)\text{ (nab)}}&=
-\frac{8\bar{z}_4^2\left(s_{34}+s_{[12]3}\right)}{s_{34}{}^2s_{[12]3}(1-z_{34})^2}
\,, \\
%%%%%%%%%
%%%%%%%%%
D_{3344,g_{(12)}\bar q_3 q_4}^{(g)\text{ (nab)}}&=
-\frac{8\bar{z}_ 3^2\left(s_{34}+s_{[12]4}\right)}{s_{34}{}^2s_{[12]4}(1-z_{34})^2}
\,, \\
%%%%%%%%%
%%%%%%%%%
E_{33,g_{(12)} \bar q_3 q_4}^{(g)\text{ (nab)}} &=
\frac{2z_4^2(s_{[12]3}+s_{34})-4z_3z_4s_{[12]4}}{(1-z_{34})z_{34}s_{34}} 
\left(\frac{1}{s_{[12]4}}+\frac{1}{s_{34}} \right) 
+\frac{2z_4^2s_{[12]3}}{(1-z_{34})z_{34}s_{34}^2}
\,, \\
%%%%%%%%%%
%%%%%%%%%%
E_{44,g_{(12)}\bar q_3 q_4}^{(g)\text{ (nab)}}&=
\frac{2z_3^2(s_{[12]4}+s_{34})-4z_3z_4s_{[12]3}}{(1-z_{34})z_{34}s_{34}} 
\left(\frac{1}{s_{[12]3}}+\frac{1}{s_{34}} \right) 
+\frac{2z_3^2s_{[12]4}}{(1-z_{34})z_{34}s_{34}^2}
\,, \\
%%%%%%%%%%
%%%%%%%%%%
E_{34,g_{(12)}\bar q_3 q_4}^{(g)\text{ (nab)}}&=
\frac{s_{34}+s_{[12]4}}{2s_{34}s_{[12]3}z_{34}}
-\frac{(s_{34}+s_{[12]3})}{2s_{34}s_{[12]4}}
\left(
\frac{4z_4\bar{z}_3}{(1-z_{34})z_{34}}
+ \frac{1}{z_{34}}
\right)
\nonumber\\
&\text{}\hspace{2.5cm}+ \frac{4\bar{z}_3}{s_{34}(1-z_{34})z_{34}}
\left(
z_3-\frac{1}{2}z_4 - \frac{s_{[12]3}z_4-s_{[12]4}z_3}{s_{34}}
\right)
\,, \\
%%%%%%%%%%
%%%%%%%%%%
E_{43,g_{(12)}\bar q_3 q_4}^{(g)\text{ (nab)}}&=
\frac{s_{34}+s_{[12]3}}{2s_{34}s_{[12]4}z_{34}}
-\frac{(s_{34}+s_{[12]4})}{2s_{34}s_{[12]3}}
\left(
\frac{4z_3\bar{z}_4}{(1-z_{34})z_{34}}
+ \frac{1}{z_{34}}
\right)
\nonumber\\
&\text{}\hspace{2.5cm}+ \frac{4\bar{z}_4}{s_{34}(1-z_{34})z_{34}}
\left(
z_4-\frac{1}{2}z_3 + \frac{s_{[12]3}z_4-s_{[12]4}z_3}{s_{34}}
\right)
\,, \\
%%%%%%%%%%
%%%%%%%%%%
F_{g_{(12)}\bar q_3 q_4}^{(g)\text{ (nab)}}&= 
\frac{1}{4}-\frac{\left(z_3s_{[12]4}-z_4s_{[12]3}\right)^2}{s_{34}{}^2z_{34}^2}+\frac{\left(s_{34}
+s_{[12]4}\right)\left(s_{34}-z_3s_{[12]4}\right)}{2s_{34}s_{[12]3}z_{34}}
-\frac{s_{34}+s_{[12]3}}{2s_{[12]4}} \nonumber \\
&+\frac{s_{[12]4}\left(z_{34}\left(4-3z_4\right)-8z_3{}^2\right)}{2s_{34}z_{34}^2}
+\frac{3\left(1-z_{34}\right)}{2z_{34}}+\frac{z_3z_4}{z_{34}^2}
+ (3\leftrightarrow 4)
\,.
\end{align}

\newpage

\bibliographystyle{JHEP}
\bibliography{quadri}
\end{document}